\documentclass[lettersize,journal]{IEEEtran}
\usepackage{amsmath,amsfonts}
\usepackage{algorithmic}
\usepackage{algorithm}
\usepackage{array}
\usepackage[caption=false,font=normalsize,labelfont=sf,textfont=sf]{subfig}
\usepackage{textcomp}
\usepackage{stfloats}
\usepackage{url}
\usepackage{verbatim}
\usepackage{graphicx}
\usepackage{cite}
\usepackage{xcolor}
\usepackage{soul}
\usepackage{hyperref}
\usepackage[utf8x]{inputenc}
\usepackage{empheq}
\usepackage[acronym]{glossaries}  % For making a list of acronyms
\usepackage{makecell}
\makeglossaries

\DeclareMathOperator*{\argmin}{arg\,min}
\newcommand{\probP}{\text{I\kern-0.15em P}}
\newcommand{\ubar}[1]{\text{\b{$#1$}}}
\allowdisplaybreaks
\begin{document}

\title{Mitigating Increase-Decrease Gaming with Alternative Connection Agreements: A Defender- Attacker-Defender Game}
% Coordinating congestion management instruments to prevent increase-decrease gaming 

\author{Bart van der Holst,~\IEEEmembership{Student Member,~IEEE, Thomas Swarts,\\~\IEEEmembership{Student Member,~IEEE}, Phuong Nguyen,\\~\IEEEmembership{Member,~IEEE}, Johan Morren,~\IEEEmembership{Member,~IEEE}, Koen Kok,~\IEEEmembership{Member,~IEEE}}
%\author{Phuong Nguyen,~\IEEEmembership{Member,~IEEE,}
% \author{Johan Morren,~\IEEEmembership{Member,~IEEE,}
% \author{Koen Kok,~\IEEEmembership{Member,~IEEE,}
        % <-this % stops a space
% \thanks{\textcolor{red}{TODO:}....}% <-this % stops a space
\thanks{The Netherlands Enterprise Agency financially supported
this work through the GO-E project (MOOI32001).}
\thanks{All authors are with the Eindhoven University of Technology, Department of Electrical Engineering, Eindhoven, 5612 AP. Johan Morren is also with Enexis Netbeheer, Asset Management, 's Hertogenbosch, 5223 MB.}}


% TODO: test

\maketitle

\begin{abstract}
% This paper presents a trilevel defender-attacker-defender model to investigate the potential of alternative connection agreements (ACAs) to prevent increase-decrease gaming in redispatch markets in distribution networks. 
Redispatch markets are widely used by system operators to manage network congestion. A well-known drawback, however, is that Flexibility Service Providers (FSPs) may strategically adjust their baselines in anticipation of redispatch actions, thereby aggravating congestion and raising system costs. To address this increase–decrease gaming, Distribution System Operators (DSOs) could use Alternative Connection Agreements (ACAs) to conditionally limit the available connection capacity of market participants in the day-ahead stage. In this paper, we present a novel Defender-Attacker-Defender game to investigate the potential of this approach in distribution networks under load and price uncertainty. We solve the resulting trilevel optimization model using a custom branch-and-bound algorithm, and we demonstrate that it efficiently solves the problem without exploring many nodes in the branch-and-bound search tree for most simulated scenarios. The case study demonstrates that applying ACAs can substantially lower redispatch costs (e.g. by 25\%) for the DSO with only a limited impact on FSP profits. The effectiveness of the approach critically depends on how often the DSO can invoke ACAs and on the extent to which the DSO can anticipate strategic bidding behavior of the FSP.

% Redispatch markets are one of the most common tools for system operators to mitigate congestion in their networks. However, a well-known concern with redispatch markets is that when Flexibility Service Providers (FSPs) can anticipate redispatch actions, their bidding behavior can increase congestion and costs for the system operator. One potential solution to prevent this increase-decrease gaming is for the Distribution System Operators (DSOs) to use alternative connection agreements (ACAs) to reduce the connection capacity of the redispatch market participants day-ahead. In this paper, we present a trilevel defender-attacker-defender model to investigate the potential of this approach in distribution networks. The trilevel model includes both the effect of load and price uncertainty, and we provide a solution strategy based on a custom Branch-and-Bound algorithm. We find that our solution strategy efficiently solves the problem without exploring many nodes for most simulated scenarios. We also find that using ACAs can significantly reduce redispatch costs for the DSO, while not decreasing much profits for FSPs at the same time. The effectiveness of applying ACAs for this purpose depends on the number of times the DSO can invoke the ACA on a daily basis, and whether the DSO can strategically anticipate the gaming strategy of the DSO.

% 150-200 wordsistribution System Operators (DSOs) adopting
\end{abstract}

\begin{IEEEkeywords}
Redispatch Markets, Increase-Decrease Gaming, Connection Agreements, Multi-level Optimization, Branch-and-Bound, Monte Carlo Simulation
\end{IEEEkeywords}

\newacronym{dso}{DSO}{Distribution System Operator}
\newacronym{tso}{TSO}{Transmission System Operator}
\newacronym{lfm}{LFM}{Local Flexibility Market}
\newacronym{fsp}{FSP}{Flexibility Service Provider}
\newacronym{da}{DA}{day-ahead}
\newacronym{mpec}{MPEC}{Mathematical Program with Equilibrium Constraints}
\newacronym{kkt}{KKT}{Karush-Kuhn-Tucker}
\newacronym{lmp}{LMP}{Locational Marginal Pricing}
\newacronym{aca}{ACA}{Alternative Connection Agreement}
\newacronym{rc}{RC}{Redispatch Contract}
\newacronym{bnb}{BnB}{Branch-and-Bound}
\newacronym{ev}{EV}{Electric Vehicles}
\newacronym{kde}{KDE}{Kernel Density Estimation}
\newacronym{mv}{MV}{medium-voltage}
\newacronym{milp}{MILP}{Mixed-Integer Linear Program}
\newacronym{cs}{CS}{complementary slackness}

\section{Introduction}
\label{sec: Introduction}
\subsection{Background and Motivation}
The rapid increase in renewable generation and the ongoing electrification of energy demand have led to increased congestion in electricity networks worldwide \cite{IEA}. While grid reinforcement has traditionally been the main approach to mitigate congestion, the most widely used alternative is \textit{redispatch}. With redispatch, the system operator procures flexibility from \glspl{fsp} between spot market closure and the actual delivery of energy. The \glspl{fsp} are then compensated for the amount of energy deviated from some predefined energy schedule (or \textit{baseline}). This baseline typically reflects the outcome of trade in the \gls{da} market or a local electricity market. 

In redispatch markets, \glspl{fsp} submit offers reflecting the costs of deviating from their baseline. Currently, many countries operate redispatch markets in transmission systems, but the Netherlands \cite{GOPACS} and the UK \cite{flexiblepower} operate (local) redispatch markets in their distribution systems too. Furthermore, \glspl{dso} in many other countries are experimenting with (local) redispatch markets through pilot projects \cite{rebenaque2023success}.

However, a well-known concern with redispatch markets is that they can provide incentives for \textit{increase-decrease gaming} \cite{holmberg2015comparison, hirth2020market}. Anticipating the \gls{dso} will solve congestion with redispatch, \glspl{fsp} face incentives to manipulate their baselines to benefit from extra redispatch income. This behavior can artificially inflate the expected congestion in the network and thereby increase the redispatch actions from the \gls{dso}. This results in higher redispatch costs for the \gls{dso} and higher revenues for the \glspl{fsp}. 
% Holmberg \textit{et al.} argue that these additional revenues from increase-decrease gaming also distort long-term incentives \cite{holmberg2015comparison}, encouraging actors to locate in congested areas or prompting unnecessary grid reinforcements by \glspl{dso} \cite{hirth2020market}.

To mitigate such gaming behavior, several countermeasures have been proposed \cite{kloters2022monitoring}. One of the proposed solutions is for \glspl{dso} to apply other congestion management instruments alongside redispatch as a complementary measure \cite{pechan2023risks}. For example, network tariffs have been suggested for this purpose \glspl{fsp}, providing \glspl{fsp} with additional price signals to reduce congestion \cite{holmberg2024inc}. Furthermore, many countries are developing \glspl{aca} as a means to procure flexibility in distribution networks \cite{CEER}. Some of these contracts allow \glspl{dso} to mitigate expected congestion by limiting the contracted connection capacity of \glspl{fsp} day-ahead, in exchange for discounts on the network tariff. In the Netherlands, several large batteries providing redispatch also have an \gls{aca} of this type. This setting raises the question of whether these \glspl{aca} could also be used as a complementary measure to prevent increase-decrease gaming in redispatch markets in distribution networks. This will be the central question of this paper.

\subsection{Related Work}
\label{sec: Related Work}

The most widely adopted modeling approach to quantify the effects of strategic bidding in redispatch markets is to use (stochastic) bilevel optimization models \cite{beckstedde2023bilevel, wang2020aggregation, farrokhseresht2020strategic, pozzi2025analysing}. In these models, the bidding problem of an \gls{fsp} is formulated as a leader-follower game, in which the upper level represents the strategic \gls{fsp}, and the lower level captures the redispatch market-clearing process. Given the lower level is convex, it can be replaced by its \gls{kkt} conditions, enabling reformulation as a single-level \gls{mpec}. These approaches typically assume that only a single market participant acts strategically, while all others behave competitively. For settings with multiple strategic agents, Nash equilibria of the redispatch market clearing problem can either be derived analytically under many simplifying assumptions \cite{marques2024strategic, silani2024stochastic}, or obtained by solving the problem as an equilibrium problem with equilibrium constraints \cite{sarfati2020simulation}. The work of Silani \textit{et al.} is an example of the former, and considers the extreme case of a local flexibility market with an infinite number of participants, adopting a stochastic mean field approach \cite{silani2024stochastic}. Similar to \cite{sarfati2020simulation}, they considered load uncertainty, but price uncertainty was not considered. Zhang \textit{et al.} and Ye \textit{et al.} study increase-decrease gaming in local flexibility markets using multi-agent reinforcement learning simulations \cite{ye2022multi, zhang2025arbitrage}. Unlike analytical and optimization-based approaches, the simulated agents lacked information about others’ network or cost parameters, yet learned increase-decrease gaming as an optimal strategy during training.

Several methods have been proposed to mitigate increase–decrease gaming in redispatch markets. Broadly, these methods can be grouped into: baseline monitoring, baselining based on historical data, imposing price caps, long-term contracts, introducing randomness, and applying other congestion management instruments \cite{kloters2022monitoring, pechan2023risks}. Since increase-decrease gaming arises from inflated self-reported baselines, various baseline monitoring strategies \cite{jahns2023prevention, zhang2025review} and methods to construct baselines based on historical data \cite{wijaya2014bias} have been explored. However, a disadvantage of these methods is that their effectiveness reduce when \glspl{fsp} construct baselines that seem reasonable given their asset portfolios.

With price gaps and long-term contracts, the \gls{dso} aims to reduce the revenues from increase-decrease gaming by capping or fixing the bid prices from the \glspl{fsp} in the redispatch market \cite{bjorndalen2020market}. These approaches limit the price spread between the \gls{da} market and redispatch market, but do not remove the incentive for increase-decrease gaming altogether. To further increase the risk of the increase-decrease gaming strategy, Vuelvas \textit{et al.} \cite{vuelvas2018limiting} and Muthirayan \textit{et al.} \cite{muthirayan2019mechanism} propose baseline mechanisms that introduce a degree of randomness in the redispatch bid selection. They show that if the degree of randomness is small enough, increase-decrease gaming can be reduced without significantly reducing market efficiency. These results hold for a large number of market participants, which is not the typical case in distribution networks. 

Finally, Pechan \textit{et al.} propose using complimentary congestion management instruments as a potential solution to increase-decrease gaming \cite{pechan2023risks}. Dubbeling \textit{et al.} similarly suggest that \glspl{aca} could be used for this purpose  \cite{dubbeling2025congestion} when discussing the status of the existing redispatch market in the Dutch distribution system, specifically. 

To the authors’ knowledge, the potential of applying \glspl{aca} to prevent increase–decrease gaming has only been proposed as a theoretical solution, but has not been quantified before. In this paper, we take the first step to address this gap by formulating the resulting decision process as a \textit{defender-attacker-defender} game. In this setting, the \gls{dso} first uses \glspl{aca} to prevent \glspl{fsp} from increase-decrease gaming (defend). The \gls{fsp} then responds by sending its redispatch baseline (attack). The \gls{dso} then relieves the expected congestion in the network using redispatch (defend). Calculating the outcomes of this game for both players under various strategies provides insight into the potential of \glspl{aca} as a solution to increase-decrease gaming in distribution networks. In contrast to existing work on increase-decrease gaming, we include both load and price uncertainty in our analysis.

\subsection{Contributions}
To better understand the potential of applying \glspl{aca} against increase-decrease gaming, this paper presents the following contributions:
\begin{itemize}
    \item Formulating a defender-attacker-defender game to study the effectiveness of \glspl{aca} against increase-decrease gaming in distribution networks. Within this game, the \gls{dso} needs to solve a novel trilevel optimization problem to determine how to exercise the \glspl{aca}.
    \item Proposing a novel solution approach for the trilevel problem by reformulating it as a bilevel problem with a non-convex integer lower-level problem, and solving it using a custom \gls{bnb} algorithm.
    \item Exploring the effect of load and price uncertainty on the potential gains/losses resulting from increase-decrease gaming in a case study.
    \item Quantifying the added value of \glspl{dso} anticipating increase–decrease gaming by \glspl{fsp}.
\end{itemize}
% The proposed framework can be used for systematic evaluation of \glspl{aca} as a measure to prevent increase-decrease gaming in distribution networks.
% The proposed framework enables systematic evaluation of \glspl{aca} as a congestion management instrument to prevent increase-decrease gaming in \glspl{lfm}. It can be used by \glspl{dso} to assess the viability of deploying \glspl{aca} in situations where strategic behavior is suspected.
% \input{Assumptions/Assumptions}
\section{The Defender-Attacker-Defender Game}
To formulate the Defender–Attacker–Defender game, we need to specify the design of the \gls{aca} and redispatch processes. We base these on the current state of the Dutch distribution system. The Netherlands provides an ideal case study because the country operates as a single bidding zone in the \gls{da} market and uses a national redispatch platform, called GOPACS~\cite{GOPACS}, which is accessible to both the Transmission System Operator TenneT and the \glspl{dso}. This setup creates opportunities for increase–decrease gaming by \glspl{fsp} active in distribution networks. Moreover, several \glspl{aca} have been introduced in the Netherlands since 2024 ~\cite{acm2024}.

\autoref{fig: actors} presents an overview of the Defender-Attacker-Defender game. We consider a \gls{mv} distribution network with a single \gls{fsp}\footnote{Currently, only a few \glspl{fsp} provide redispatch services in Dutch distribution networks, and the likelihood of multiple \glspl{fsp} being active within a single \gls{mv} network remains low. Our model can be extended to include multiple \glspl{fsp}, but only if the added \glspl{fsp} do not anticipate future redispatch income.}. The game between the \gls{dso} and this \gls{fsp} always consists of three steps. As a first step, the \gls{dso} can exercise the \gls{aca} of the \gls{fsp} (Defend). We model this \gls{aca} after the Dutch ATR85/15 connection agreement~\cite{acm2024}. The ATR85/15 is currently only available in the transmission system, but \glspl{dso} are investigating its implementation in distribution systems too~\cite{acm2024}. With this \gls{aca}, the \gls{fsp} has firm connection capacity $\bar{p}$ for 85\% of the hours of the year. For the remaining 15\% of the time, the \gls{dso} can reduce the \gls{fsp}'s capacity on its connections to some lower value $\bar{p}^{\mathrm{ACA}}$ specified in the contract. Exercising the contract takes place on the day prior to delivery, before the \gls{da} market closes. To determine how to apply the \gls{aca}, we compare two possible strategies (see \autoref{fig: actors}): one in which the \gls{dso} does not anticipate increase–decrease gaming (\textit{ACA No Anticipation}), and one in which it does (\textit{ACA Anticipation}).

Immediately after \gls{aca} activation, the \gls{fsp} communicates its baseline for the following day (Attack). Here, we again distinguish between two strategies: one in which the \gls{fsp} cannot anticipate future redispatch income, and can thus not engage in increase–decrease gaming (\textit{No Gaming}), and one in which it can and adjusts its baseline accordingly (\textit{Gaming}).

In the third step, the \gls{dso} applies redispatch to prevent congestion in the network. Since there is only a single \gls{fsp} in the network, redispatch is procured through a \gls{rc}. The \gls{rc} was introduced in the Netherlands in 2022 \cite{acm2022} and allows the \gls{dso} and \gls{fsp} to agree on a price per MWh $\pi^{\mathrm{RC}}$ for the redispatched energy. When the \gls{dso} exercises the \gls{rc}, the \gls{fsp} submits a buy/sell order to GOPACS. The market then matches the offer from the \gls{fsp} with a counter-order submitted by a market party outside the congested area. If there is a price difference $\Delta \pi^{\mathrm{RC}}$ between the two orders blocking the trade, the \gls{dso} pays this \textit{spread} to the selling party. For a more detailed account of how GOPACS works, we refer the reader to \cite{van2024activation}. In principle, the \gls{dso} can use the \glspl{rc} to continuously request redispatch intraday, up till 45 minutes before real time. However, doing so across multiple distribution networks is a highly complex operational challenge for \glspl{dso}. For this reason, we assume that the \gls{dso} applies the \gls{rc} intraday only once, shortly before delivery. We thus assume that all uncertain parameters have realized at this point. We consider only a single strategy for this decision (\textit{Redispatch}).

We also make the following additional assumptions:
\begin{enumerate}
    \item The \gls{fsp} manages a portfolio consisting solely of \glspl{ev}\footnote{Several \gls{ev} aggregators are already active in the Netherlands. The framework can be extended to heterogeneous \gls{fsp} portfolios, provided asset constraints can be formulated in a mixed-integer linear form.}. 
    \item \label{ass: ev uncertainty} The future charging behavior of the \gls{ev} fleet is assumed to be known to the \gls{fsp}. 
    %This work focuses on uncertainty from the \gls{da} price and the uncontrolled loads in the network, as formalized in assumptions~\ref{ass: load uncertainty} and~\ref{ass: price uncertainty}.
    \item \label{ass: load uncertainty} Generation and loads at connections outside the  \gls{fsp} portfolio are considered to be inflexible and uncertain.
    \item \label{ass: price uncertainty} The \gls{fsp} is also a Balance Responsible Party and buys its electricity on the \gls{da} market under \gls{da} price uncertainty. The \gls{fsp} is a price taker in this market.
    \item \label{ass: DA=BL} The redispatch baseline communicated by the \gls{fsp} to the \gls{dso} matches the \gls{da} market schedule of the \gls{fsp}. Note that if the \gls{fsp} applies increase-decrease gaming, the \gls{fsp} anticipates that its \gls{da} market schedule will be different from the final dispatch.     
    \item  \label{ass: ev constraints} To avoid detection of increase-decrease gaming by the \gls{dso}, the \gls{fsp} ensures that the redispatch baseline is feasible given the \gls{ev} asset constraints.
    \item The \gls{fsp} does not deviate from the \gls{da} schedule/baseline unless it provides redispatch. The \gls{fsp} is thus not active on intraday or balancing markets.
    \item \label{ass: counter order} We assume that there is always a counter-order available for the \gls{fsp}'s redispatch order. The price of this counter-order is uncertain on the day before delivery. 
    \item \label{ass: complete information} The \gls{dso} and \gls{fsp} have complete information about eachother's objectives and parameters. 
    %This assumption is required for the players to anticipate each other's actions.
\end{enumerate}

Assumptions ~\ref{ass: ev uncertainty} and~\ref{ass: complete information} are arguably the most significant simplifications compared to real-world practice. However, assuming complete knowledge of the \gls{ev} charging profile focuses the analysis on external sources of uncertainty: 1) the prices in the \gls{da} market; 2) the prices of counter-orders in GOPACS, which determine the spread paid by the \gls{dso}; and 3) the loads at connections outside the \gls{fsp} portfolio. These parameters influence increase-decrease gaming behavior regardless of the specific assets in the \gls{fsp}'s portfolio. These three parameters are considered to be unknown during the first two steps of the game (see \autoref{fig: actors}). Moreover, the assumption of complete information is widely used in game-theoretic studies, as it enables a tractable analysis of strategic interactions and the computation of equilibrium outcomes. In this context, the model may overestimate both the \gls{dso}’s ability to strategically activate the \gls{aca} and the \gls{fsp}’s potential for gaming. The results can thus be interpreted as upper bounds on the performance of these strategies. 

% Je gebruikt nu decision process en game door elkaar.
% \input{Gaming/Gaming}
\section{Decision Models}
\label{sec: Model Formulation}
% zie https://ieeexplore.ieee.org/stamp/stamp.jsp?tp=&arnumber=7317827
% Assumptions, nomenclature
\begin{figure}[!t]
\centering
\includegraphics[width=0.9\linewidth]{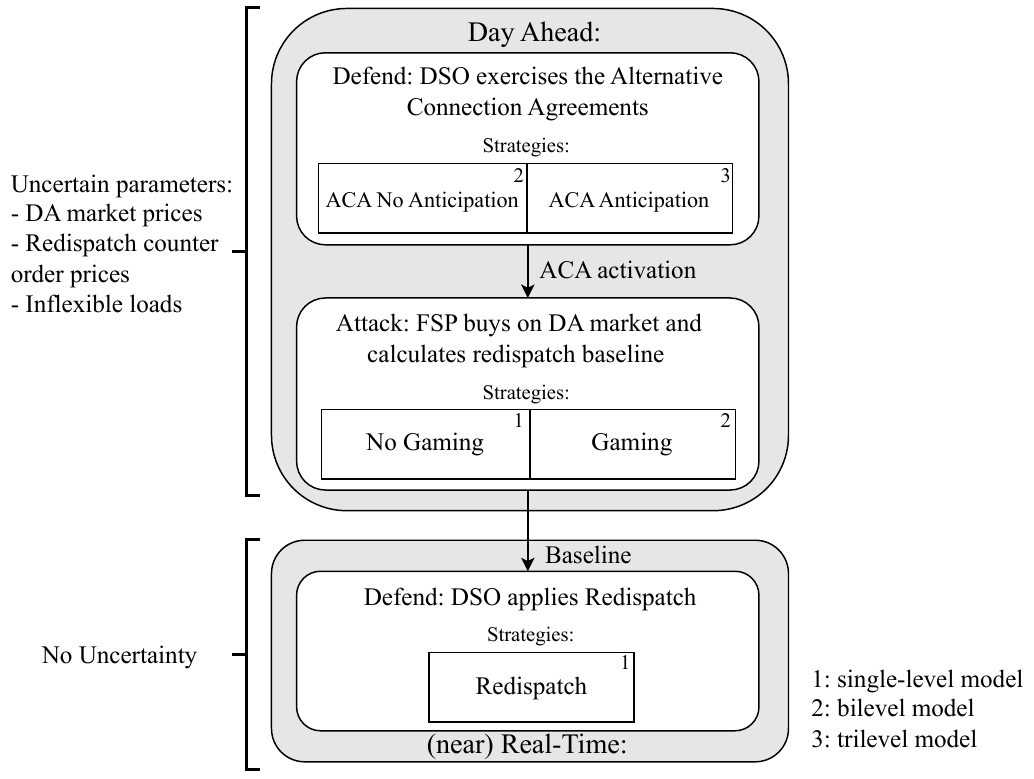}
\caption{Diagram of the Defender–Attacker–Defender game, where the \gls{dso} exercises \glspl{aca} as a preventive measure against increase–decrease gaming. The game consists of three sequential decisions: for the first two, we compare two alternative strategies, while for the final decision, only one strategy is considered. Each strategy is represented by a corresponding single- or multi-level optimization problem (see Section \ref{sec: Other problems}).}
% \caption{Diagram of the decision process for preventing increase-decrease gaming with alternative connection agreements, following from the assumptions from Section \ref{sec: Assumptions}. Decisions taken during the day-ahead stage involve uncertainty about \gls{da} prices, redispatch price spreads, and inflexible loads. For both the activation of the alternative connection agreement by the \gls{dso} and for the \gls{fsp} creating the baseline, two different strategies are considered.}
\label{fig: actors}
\end{figure}

The Defender-Attacker-Defender game presented in the previous section consists of three stages. At every stage, either the \gls{dso} or the \gls{fsp} takes a decision by solving an optimization problem. This section presents these optimization problems in detail. There are five of these models in total: two possible strategies for the \gls{dso} applying the \gls{aca} (\textit{ACA No Anticipation} and \textit{ACA Anticipation}), two strategies for the baseline communication stage by the \gls{fsp} (\textit{No Gaming} and \textit{Gaming}), and only one \textit{Redispatch} strategy (see \autoref{fig: actors}). In the next subsection, we will discuss the trilevel optimization problem the \gls{dso} solves when applying the \textit{ACA Anticipation} strategy. After, we discuss how the other decision problems can be viewed as reductions compared to this general setup. 
\subsection{The Trilevel model for the ACA Anticipation Strategy}
\label{sec: Decision Model}

% \autoref{fig: actors} presents the decision process between the \gls{dso} and \gls{fsp} resulting from the assumptions presented in Section \ref{sec: Assumptions}. In the day-ahead stage, the \gls{dso} first decides on when to use the \gls{aca}. This information is then communicated to the \gls{fsp}. The \gls{fsp} then buys energy on the \gls{da} market, and takes this schedule to be its baseline for redispatch (see Assumption \ref{ass: DA=BL}). To make this decision, the \gls{fsp} two strategies are considered and compared: 1) not applying increase-decrease gaming (focusing only on \gls{da} market prices), and 2) applying increase-decrease gaming. For the \gls{dso}, the \gls{aca} can be activated while anticipating the \gls{fsp} will apply increase-decrease gaming, or not. Both decisions in the day-ahead stages are taken under uncertain \gls{da} prices and loads from other nodes in the network.

% Given the \gls{dso} uses the \gls{rc} only once (see Assumption \ref{ass: 1 RC}), there is only one decision stage for redispatch, taking place close to delivery. For this decision, the realized \gls{da} prices and loads are considered. 

% Considering the two different strategies in the first two decisions, there are five decision problems in this decision process. However, the trilevel defender-attacker-defender problem that is solved by the \gls{dso} for \gls{aca} activation while considering increase-decrease gaming by the \gls{fsp}, is the most generic and will now be discussed in detail. 
\begin{figure}[!b]
\vspace{-0.4cm}
\centering
\includegraphics[width=0.7\linewidth]{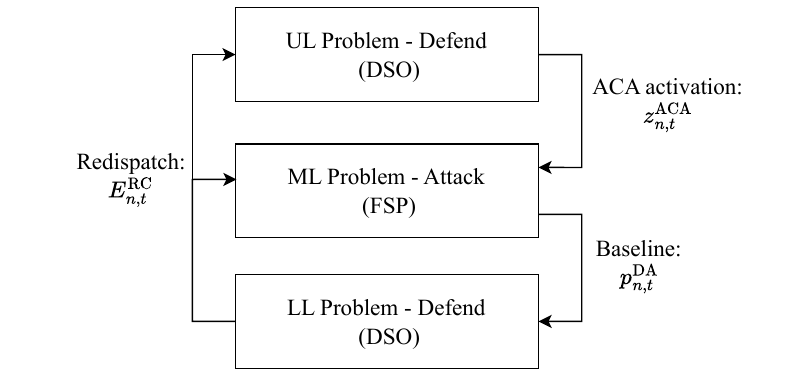}
\caption{Structure of the proposed trilevel defender-attacker-defender model. Variables propagating downwards are treated as parameters in the lower levels. Variables propagating upwards are the result of lower-level decisions, providing feedback to higher levels.}
\label{fig: trilevel}
\end{figure}

Let $t \in \Omega_T$ denote the time steps in the next day (hourly resolution $\Delta t$), let $n \in \Omega_N$ denote the nodes in the \gls{mv} network, let $(m, n) \in \Omega_B$ denote all branches in the network (cables and transformers), where $m, n \in \Omega_N$. Also, let $\Omega_{\mathrm{FSP}} \subset \Omega_N$ denote the set of all nodes in the network where the \gls{fsp} is active, and let $\omega \in \Omega_S$ denote the set of scenarios for the uncertain parameters (e.g. \gls{da} market prices). We will present the deterministic model for a given scenario $\omega$. The general trilevel structure of the problem consists of an upper level (UL), middle level (ML), and lower level (LL). Every level has an objective function $F$ and a constraint set $\mathcal{C}$. The trilevel problem is presented schematically in \autoref{fig: trilevel} and can be described by:
% Upper-level
\begin{align}
  \min_{z_{n, t}^{\mathrm{ACA}} \in \mathcal{C}_{\mathrm{UL}}} \quad & F_{\mathrm{UL}}(E^{\mathrm{RC}}_{n, t}) \label{eq:upper-level} \\
  \text{s.t. } & E^{\mathrm{RC}}_{n, t} \in \argmin_{p^{\mathrm{DA}}_{n, t} \in \mathcal{C}_{\mathrm{ML}}(z_{n, t}^{\mathrm{ACA}})} F_{\mathrm{ML}}(E^{\mathrm{RC}}_{n, t}, p^{\mathrm{DA}}_{n, t}) \label{eq:middle-level} \\
  & \text{s.t. } E^{\mathrm{RC}}_{n, t} \in \argmin_{E^{\mathrm{RC}}_{n, t} \in \mathcal{C}_{\mathrm{LL}}(z_{n, t}^{\mathrm{ACA}}, p^{\mathrm{DA}}_{n, t})} F_{\mathrm{LL}}(E^{\mathrm{RC}}_{n, t}), \label{eq:lower-level}
\end{align}
where $n\in \Omega_{\mathrm{FSP}}$ and $t \in \Omega_T$. the variables $z^{\mathrm{ACA}}_{n, t}$ denote the binary \gls{aca} activations, $p^{\mathrm{DA}}_{n, t}$ denote the active power baseline, and $E^{\mathrm{RC}}_{n, t}$ denote the requested redispatch volumes through the \gls{rc}. The indents are used to illustrate that e.g. solving the ML problem, requires both ML feasibility $p^{\mathrm{DA}}_{n, t} \in \mathcal{C}_{\mathrm{ML}}(z_{n, t}^{\mathrm{ACA}})$ and LL optimality \eqref{eq:lower-level}. We now discuss the three levels in more detail.
\subsubsection{Upper Level- Exercising ACAs}
In the UL, the \gls{dso} solves for the optimal \gls{aca} activation $z_{n,t}^{\mathrm{ACA}}$ for all connections of the \gls{fsp}. The objective function is a function of the requested redispatch volumes $E^{\mathrm{RC}}_{n, t}$ as the \gls{dso} minimizes its redispatch costs for the next day:
\begin{equation}
    F_{\mathrm{UL}}(E^{\mathrm{RC}}_{n, t}) = \sum_{n \in \Omega_{\mathrm{FSP}}} \sum_{t \in \Omega_T} \Delta \pi_{t, \omega}^{\mathrm{RC}} E^{\mathrm{RC}}_{n, t},
\end{equation}
where the \gls{dso} pays the spread price between the order of the \gls{fsp} at price level $\pi^{\mathrm{RC}}$ and some counter in GOPACS (see Assumption \ref{ass: counter order}). This price spread $\Delta \pi_{t, \omega}^{\mathrm{RC}}$ is unknown at this decision stage, and thus depends on the scenario $\omega$. The constraint set $\mathcal{C}_{\mathrm{UL}}$ is simply:
\begin{align}
    & z_{n,t}^{\mathrm{ACA}} \in \{0, 1\},  && n \in \Omega_{\mathrm{FSP}}, t \in \Omega_T, \\
    & \sum_{t\in \Omega_T} z_{n,t}^{\mathrm{ACA}} \leq B^{\mathrm{ACA}}, && n \in \Omega_{\mathrm{FSP}},
\end{align}
where $B^{\mathrm{ACA}}$ is the number of times the \gls{dso} can exercise the \gls{aca} (i.e. a daily budget). In principle, the \gls{dso} can use the \gls{aca} as many times as it wants on a daily basis, as long as it does not exceed using it for more than 15\% of the hours in the year. However, by adding the budget constraint and the budget as a parameter, we can investigate the added value of applying the \gls{aca} less or more frequently within a day.
\subsubsection{Middle Level - Constructing the Baseline}
\begin{figure}[!t]
\centering
\includegraphics[width=0.8 \linewidth]{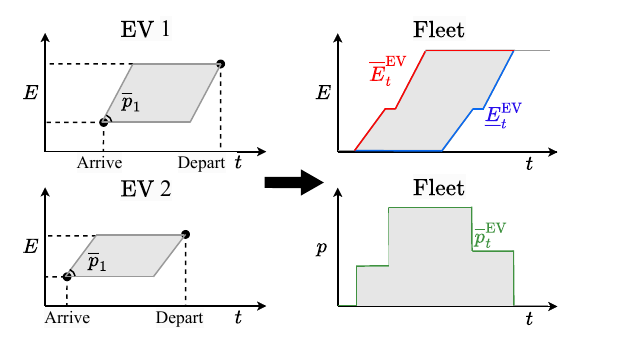}
\caption{Illustration of how the parameters $\bar{E}^{\mathrm{EV}}$, $\ubar{E}^{\mathrm{EV}}$, and $\bar{p}^{\mathrm{EV}}$ can be calculated for a fleet of two cars from their respective charging session information. The two figures in the left column present the areas containing all possible charging schedules of the individual EVs. These areas are characterized by the maximum charge power of the respective EVs, and the arrival/departure times and state of charge. In the right column, the figures present the three fleet parameters based on the charging sessions.}
\label{fig: EVs}
\end{figure}

In the ML, the \gls{fsp} solves for the optimal baseline $p_{n,t}^{\mathrm{DA}}$ for all connections in its portfolio. In the case the \gls{fsp} constructs its baseline strategically, considering both \gls{da} costs and future redispatch income, its objective function is:
\begin{equation}
        F_{\mathrm{ML}}(E^{\mathrm{RC}}_{n, t}, p_{n,t}^{\mathrm{DA}}) = \sum_{n \in \Omega_{\mathrm{FSP}}} \sum_{t \in \Omega_T} \left(-\pi^{\mathrm{RC}} E^{\mathrm{RC}}_{n, t} + \pi_{t, \omega}^{\mathrm{DA}} p^{\mathrm{DA}}_{n, t} \Delta t \right) \label{eq: F ML},
\end{equation}
\begin{equation}
        \min_{p_{n, t}^{\mathrm{DA}}} \sum_{n \in \Omega_{\mathrm{FSP}}} \sum_{t \in \Omega_T} \pi_{t, \omega}^{\mathrm{DA}} p^{\mathrm{DA}}_{n, t} \Delta t  \label{eq: F ML},
\end{equation}
\begin{equation}
        \min_{z_{n, t}^{\mathrm{ACA}}} \sum_{n \in \Omega_{\mathrm{FSP}}} \sum_{t\in\Omega_T} \Delta \pi_{t, \omega}^{\mathrm{RC}} E^{\mathrm{RC}}_{n, t} \label{eq: F ML},
\end{equation}

where \gls{da} prices $\pi^{\mathrm{DA}}_{t, \omega}$ are uncertain and the redispatch price $\pi^{\mathrm{RC}}$ is specified in the \gls{rc}. The constraint set $\mathcal{C}_{\mathrm{ML}}(z_{n,t}^{\mathrm{ACA}})$ is parameterized by the \gls{aca} activation from the UL and is given by:
\begin{align}
     & 0 \leq p^{\mathrm{DA}}_{n, t}\leq \bar{p}_{n} + (\bar{p}^{\mathrm{ACA}}_{n} - \bar{p}_{n})z^{\mathrm{ACA}}_{n, t},\label{eq: aca}  && n \in \Omega_{\mathrm{FSP}}, t \in \Omega_T, \\
     & 0 \leq  p^{\mathrm{DA}}_{n, t}\leq \bar{p}^{\mathrm{EV}}_{n, t}, \label{eq: ev pmax}  && n \in \Omega_{\mathrm{FSP}}, t \in \Omega_T, \\
     &\bar{E}^{\mathrm{EV}}_{n, t} \leq \sum_{\tau\leq t} p^{\mathrm{DA}}_{n, \tau} \Delta t\leq \ubar{E}^{\mathrm{EV}}_{n, t}, \label{eq: ev soc}  && n \in \Omega_{\mathrm{FSP}}, t \in \Omega_T.
\end{align}
This first equation \eqref{eq: aca} describes how the \gls{aca} impacts the available connection capacity on every node the \gls{fsp} is active. The equations \eqref{eq: ev pmax} and \eqref{eq: ev soc} guarantee that the baseline satisfies the asset constraints for the \gls{ev} fleets (see Assumption \ref{ass: ev constraints}). To model the \gls{ev} fleets, we adopt an aggregated virtual battery model from \cite{tang2016aggregated}. The model aggregates data of charging sessions from individual \glspl{ev} (arrival/departure times and state of charge) into three fleet parameters. These parameters are: the cumulative energy charged under \textit{fast-as-possible} charging operation $\bar{E}^{\mathrm{EV}}_{n, t}$, the cumulative energy charged under \textit{slow-as-possible} operation $\ubar{E}^{\mathrm{EV}}_{n, t}$, and the instantaneous maximum charging power of the fleet $\bar{p}^{\mathrm{EV}}_{n, t}$. Fig. \ref{fig: EVs} demonstrates how these parameters can be calculated for a fleet consisting of two \glspl{ev}, but for further details we refer the reader to \cite{tang2016aggregated}. For calculating these fleet parameters, we use the open data set of domestic charging sessions in the UK from 2017 \cite{evdata}.

\subsubsection{Lower Level - Applying Redispatch}
In the LL, the \gls{dso} uses redispatch $E^{\mathrm{RC}}_{n, t}$ to solve the predicted congestion in the network based on the baselines from the \gls{fsp} and the predicted loads at other nodes in the network. The objective function of the \gls{dso} in the LL is the same as in the UL:
\begin{equation}
% \vspace{-2cm}
        F_{\mathrm{LL}}(E^{\mathrm{RC}}_{n, t}) = \sum_{n \in \Omega_{\mathrm{FSP}}} \sum_{t\in\Omega_T} \Delta \pi_{t, \omega}^{\mathrm{RC}} E^{\mathrm{RC}}_{n, t} \label{eq: F LL}.
% \vspace{-2cm}
\end{equation}
 The constraint set $\mathcal{C}_{\mathrm{LL}}(z_{n,t}^{\mathrm{ACA}}, p_{n,t}^{\mathrm{DA}})$ is parameterized by the \gls{aca} activation and baselines from the UL and ML, respectively. It is described by the following constraints:
  \begingroup
\allowdisplaybreaks
\begin{align}
     & E^{\mathrm{RC}}_{n, t} \geq 0, \label{eq: redispatch 0}  &&  n \in \Omega_{\mathrm{FSP}}, t \in \Omega_T, \\
     & p^{\mathrm{EV}}_{n, t}\Delta t = p^{\mathrm{DA}}_{n, t}\Delta t - E^{\mathrm{RC}}_{n, t}, \label{eq: redispatch} && n \in \Omega_{\mathrm{FSP}}, t \in \Omega_T, \\ 
     & p_{n, t} = p^{\mathrm{IF}}_{n, t, \omega} + \delta^{\mathrm{EV}}_{n} p^{\mathrm{EV}}_{n, t}, \label{eq: p balance} && n \in \Omega_N, t \in \Omega_T, \\
      & q_{n, t} = \lambda^{\mathrm{IF}}_{n} p^{\mathrm{IF}}_{n, t, \omega} + \delta^{\mathrm{EV}}_{n} \lambda^{\mathrm{EV}} p^{\mathrm{EV}}_{n, t}, \label{eq: q balance} && n \in \Omega_N, t \in \Omega_T, \\
     & \sum_{(m, n)} P_{(m, n), t} - \sum_{(n, o)} P_{(n, o), t} = p_{n, t}, \label{eq: P balance} && n  \in \Omega_N, t \in \Omega_T,\\
      & \sum_{(m, n)} Q_{(m, n), t} - \sum_{(n, o)} Q_{(n, o), t} = q_{n, t}, \label{eq: Q balance} && n  \in \Omega_N, t \in \Omega_T,\\
    & v_{n,t}^{2} = v_{m, t}^{2} + 2\big(R_{(m, n)} P_{(m, n),t}\nonumber \\
&\quad\quad + X_{(m, n)}Q_{(m, n),t}\big), \label{eq: V2} && (m, n) \in \Omega_B, t \in \Omega_T, \\
      % & v_{r,t}^{2} = V^{2} \label{eq: V2 root} && t \in \Omega_T\\
      & \ubar{v}^{2} \leq v_{n, t}^{2}\leq \bar{v}^{2}, \label{eq: V2 lb ub} && n  \in \Omega_N, t \in \Omega_T,\\
     & P_{(m, n), t}^2 + Q_{(m, n), t}^2 \leq \bar{S}_{(m, n)}^2, \label{eq: S ub} && (m, n) \in \Omega_B, t \in \Omega_T, \\
      & 0 \leq p^{\mathrm{EV}}_{n, t}\leq \bar{p}_{n} + (\bar{p}^{\mathrm{ACA}}_{n} - \bar{p}_{n})z^{\mathrm{ACA}}_{n, t},\label{eq: aca 2}  && n \in \Omega_{\mathrm{FSP}}, t \in \Omega_T, \\
     & 0 \leq  p^{\mathrm{EV}}_{n, t}\leq \bar{p}^{\mathrm{EV}}_{n, t}, \label{eq: ev pmax 2}  && n \in \Omega_{\mathrm{FSP}}, t \in \Omega_T \\
     &\bar{E}^{\mathrm{EV}}_{n, t} \leq \sum_{\tau\leq t}  p^{\mathrm{EV}}_{n, \tau} \Delta t\leq \ubar{E}^{\mathrm{EV}}_{n, t}, \label{eq: ev soc 2}  && n \in \Omega_{\mathrm{FSP}}, t \in \Omega_T.
\end{align}
\endgroup
Equation \eqref{eq: redispatch} relates the baseline of the \gls{fsp} to the final dispatch of the \gls{ev} fleets $p^{\mathrm{EV}}_{n, t}$ after redispatch. Equations \eqref{eq: p balance} and \eqref{eq: q balance} give the total active power and reactive power injection at every node of the network. The active power is the sum of an inflexible baseline $p^{\mathrm{IF}}_{n, t, \omega}$, which is uncertain, and the load of an \gls{ev} fleet if the \gls{fsp} is active on that node. We use the kronecker delta notation for this purpose where $\delta^{\mathrm{EV}}_{n} = 1$ if $n \in \Omega_{\mathrm{FSP}}$ and $\delta^{\mathrm{EV}}_{n} = 0$ otherwise. The factors $\lambda$ are used to compute reactive power from active power injection based on the constant power factor of the connection. Equations \eqref{eq: P balance}-\eqref{eq: S ub} are the LinDistFlow equations to describe the power flows and voltages in the radially operated \gls{mv} network (see Section \ref{sec: Results}).  Notice that equations \eqref{eq: V2} and \eqref{eq: V2 lb ub} can be considered linear when considering $v_{n, t}^2$, rather than $v_{n, t}$, as the decision variables. The parameter $\bar{S}_{(m, n)}$ denotes the apparent power limit of the branches. Finally, the constraints \eqref{eq: aca 2}-\eqref{eq: ev soc 2} mirror the ML constraints \eqref{eq: aca}-\eqref{eq: ev soc} to ensure feasible operation after redispatch actions.

The problem given by \eqref{eq:upper-level}-\eqref{eq: ev soc 2} is a mixed-integer trilevel optimization problem, with a second-order cone constraint (equation \eqref{eq: S ub}). Section \ref{sec: Solution Method} describes the solution method adopted to solve this problem. 

\subsection{The Other Decision Problems}
\label{sec: Other problems}
As mentioned at the beginning of Section \ref{sec: Model Formulation}, our analysis involves five decision problems (see also \autoref{fig: actors}). Given the formulation of the general trilevel problem for the \textit{\gls{aca} Anticipation} strategy, \autoref{tab: Strategies} presents how the optimization models for the other decision problems can be obtained.
\begin{comment}

the five decision problems can be easily obtained:
\begin{itemize}
    \item \gls{dso} exercising the \glspl{aca} while anticipating gaming (\textit{ACA Anticipation}): solve the trilevel problem.
    \item \gls{dso} exercising the \glspl{aca} while not anticipating gaming (\textit{ACA No Anticipation}): solve bilevel problem consisting of the UL and ML, while removing the redispatch term in $F_{\mathrm{ML}}$ (equation \eqref{eq: F ML}), and adding all constraints in $\mathcal{C}_{\mathrm{LL}}$ to  $\mathcal{C}_{\mathrm{UL}}$
    \item \gls{fsp} determining its baseline given \gls{aca} activation, while anticipating redispatch income (\textit{Gaming}): solve bilevel problem consisting of the ML and LL for the communicated \gls{aca} activations.
    \item \gls{fsp} determining its baseline given \gls{aca} activation, while not anticipating redispatch income (\textit{No Gaming}): solve single level problem ML, while removing redispatch term in $F_{\mathrm{ML}}$.
    \item \gls{dso} applying redispatch, given \gls{fsp} baseline (\textit{Redispatch}): solve single level LL for given baselines and \gls{aca} activations.
\end{itemize}
    
\end{comment}
 \begin{table}[h!]
  \centering
  \caption{Overview of Players' strategies and the structure of the decision model }
  \label{tab: Strategies}
   \scriptsize
  \begin{tabular}{l|l|l}
     Strategy & Description & Model\\ 
    \hline 
    \gls{aca} Anticipation & \makecell[l]{\gls{dso} exercising ACA,\\ anticipating gaming from \\ \gls{fsp}} & Trilevel problem\\
    \gls{aca} No Anticipation & \makecell[l]{\gls{dso} exercising ACA,\\ not anticipating gaming \\ from \gls{fsp}} & \makecell[l]{Bilevel problem UL-ML,\\
     without redispatch term \\
    in $F_{\mathrm{ML}}$, adding all \\ constraints from $\mathcal{C}_{\mathrm{LL}}$ to  $\mathcal{C}_{\mathrm{UL}}$} \\
    Gaming & \makecell[l]{\gls{fsp} calculates baseline, \\ while anticipating \\ redispatch income} & \makecell[l]{Bilevel problem  ML-LL\\ for given \gls{aca} activations.}\\
    No Gaming & \makecell[l]{\gls{fsp} calculates baseline, \\without anticipating \\ redispatch income} & \makecell[l]{Single-level problem ML,\\  without redispatch term\\ in $F_{\mathrm{ML}}$ for given \gls{aca} \\ activations.}\\
    Redispatch & \makecell[l]{\gls{dso} applying redispatch \\ near real time} & \makecell[l]{Single-level LL for \\ given redispatch baselines \\ and \gls{aca} activations.}\\
    % \makecell[l]{Max Possible \\ Nodes} & 2 & 48 & 1.104 & 24.288 & 510.048 & 10.200.960\\
    % Anticipation CSP & 1919.73 & 1873.95 & 1920.84 & 1882.62 & 1871.08 & 1847.83\\
    % Anticipation DSO [\%] & 1614.95 & 1332.89 & 1296.17 & 1115.07 & 1009.73 & 918.66\\
    % No Anticipation CSP & 1919.73 & 1927.95 & 1920.63 & 1919.92 & 1917.43 & 1915.94\\
    % No Anticipation DSO [\%] & 1614.95 & 1501.67 & 1297.40 & 1220.73 & 1194.00 & 1171.00\\
    % \makecell[l]{ACA No [\%] \\Anticipation } & 0.0 & 11.2 & 19.7 & 24.4 & 26.1 & 27.5\\
    % \makecell[l]{ACA [\%]\\ Anticipation } & 0.0 & 17.5 & 24.8 & 31.0 & 37.5 & 43.1\\
    \hline
  \end{tabular}
\end{table}

\begin{comment}
\begin{align}
  \min_{x \in \mathcal{C}_{\mathrm{UL}}} \quad & F(x) \label{eq:upper-level} \\
  \text{s.t. } & x \in \argmin_{y \in \mathcal{C}_{\mathrm{ML}}(x)} G(x, y) \label{eq:middle-level} \\
  & \text{s.t. } y \in \argmin_{z \in \mathcal{C}_{\mathrm{UL}}(x, y)} H(x, y, z) \label{eq:lower-level}
\end{align}
\end{comment}

\section{Uncertainty Modeling}
Three parameters are considered uncertain in the presented model. These are: the \gls{da} prices $\pi^{\mathrm{DA}}_{t, \omega}$, the redispatch price spread paid by the \gls{dso} $\Delta \pi^{\mathrm{RC}}_{t, \omega}$, and the inflexible loads on the nodes $p^{\mathrm{IF}}_{n, t, \omega}$. In this paper, we model the probability distributions of these parameters with a combined scenario set $\Omega_S$ and we then solve the defender-attacker-defender game for every scenario $\omega \in \Omega_S$ with a Monte Carlo simulation. Consequently, \gls{dso} and \gls{fsp} act as if each scenario $\omega$ were a deterministic forecast of the future, making them unaware of other possible outcomes. This approach allows us to investigate the effects of increase-decrease gaming without assuming risk preferences of the decision makers, and allows parallel computation of many smaller instances of the trilevel problem \eqref{eq:upper-level}-\eqref{eq: ev soc 2}. We first discuss the scenario generation methods for the three different uncertain parameters. Then, we elaborate on the Monte Carlo approach. 

\subsection{Scenario Generation}
\label{sec: Uncertainty Modeling}
For the \gls{da} price scenarios, we adopt a scenario generation technique from Van Wijngaarden \textit{et al.} \cite{van2024day}, based on conditioned elliptical copulas. For the redispatch price spread, we use the scenario generation method developed in our previous work, which we based on real orderbook data from GOPACS between 2023 and 2024 \cite{vanderholst2025risk}. Lastly, for the load and generation in the network, we take a simplified error-based approach, using the load profiles $\hat{p}^{\mathrm{IF}}_{n, t}$ provided with the CIGRE medium voltage distribution network (see Section \ref{sec: Results}). This network contains power profiles for industrial and residential loads, and wind and solar generation. Given some non-negative parameter $\sigma$, we define the scenarios for loads and wind generation to be sampled from:
\begin{align}
    & p^{\mathrm{IF}}_{n, t} = \hat{p}^{\mathrm{IF}}_{n, t}(1 + \epsilon_{n, t}) \label{eq: wind and load}\\
    & \epsilon_{n, t + 1} = \epsilon_{n, t} + \eta_{n, t} \label{eq: random walk} \quad \quad \eta_{n, t} \sim \mathcal{N}(0, \sigma)
\end{align}
and for solar generation from:
\begin{equation}
    p^{\mathrm{IF}}_{n, t} = \hat{p}^{\mathrm{IF}}_{n, t}(1 + \epsilon_{n, t}) \label{eq: solar} \quad \quad \epsilon_{n, t} \sim \mathcal{N}(0, \sigma).
\end{equation}
In both equations \eqref{eq: wind and load} and \eqref{eq: solar}, the uncertainty scales with the power level. However, only for loads and wind energy, the uncertainty increases as for later times due to the autoregressive relation \eqref{eq: random walk}. We can use the parameter $\sigma$ to vary the effect of load uncertainty. 
\subsection{Monte Carlo Approach}
\label{sec: Monte-Carlo}
Out of the scenario sets obtained for the \gls{da} prices, the redispatch price spread, and the loads, we construct a single scenario set $\Omega_S$ by assuming the three processes are independent. We then solve the decision process from \autoref{fig: actors} for various strategies for every scenario $\omega \in \Omega_S$ as a Monte Carlo simulation.

%To analyze the potential of \gls{aca} against increase-decrease gaming under uncertainty, the decision process presented in \autoref{fig: actors} is solved for every scenario $\omega \in \Omega_S$ as a Monte-Carlo simulation. Effectively, we thus model the \gls{dso} and \gls{fsp} as actors that are not risk-aware, acting as if the scenario is their point forecast of the future value of the parameter. This approach allows us to investigate the effects of increase-decrease gaming without assuming risk preferences or beliefs of the decision makers, and allows parallel computation of many smaller instances of the trilevel problem \eqref{eq:upper-level}-\eqref{eq: ev soc 2}, rather than solving a very large intractable problem.

To determine the number of required scenarios $N$, we apply the Gaussian stopping rule \cite{gilman1968brief}. The Gaussian stopping rule is based on the Central Limit Theorem and states that while sampling a random variable $X$ and calculating the sample mean $\bar{X}$ and sample variance $s^2$ from $N$ samples, that if:
\begin{equation}
    (1 - \alpha) - 2 \Phi\left(-\frac{\sqrt{N}\delta}{\sqrt{s^2}}\right) \geq 0 \label{eq: Gaussian stopping rule},
\end{equation}
the probability that $\bar{X}$ deviates more than $\delta$ from the true mean of the distribution of $X$, is smaller than the confidence level $\alpha$. In this inequality, $\Phi$ denotes the cumulative distribution function of the normal distribution. 

We use this principle by repeatedly solving the problem on scenarios $\omega$ untill \eqref{eq: Gaussian stopping rule} is satisfied for the objective values $F_{\mathrm{UL}}$, $F_{\mathrm{ML}}$, and $F_{\mathrm{LL}}$ for $\alpha=0.05$ and a $\delta = 0.05 s$. This typically took $10^2-10^3$ samples per simulation, depending on the decision problem and problem parameters.     
% Of lostrekken
\section{Solution Method}
\label{sec: Solution Method}
We now provide the most important steps of solving the mixed-integer trilevel problem with second-order cone constraints given by \eqref{eq:upper-level}-\eqref{eq: ev soc 2}. Our solution strategy boils down to:
\begin{enumerate}
    \item Linearizing equation \eqref{eq: S ub} with a set of linear constraints;
    \item Converting the resulting problem into a mixed-integer linear bilevel problem with a new \gls{milp} lower level $\mathrm{LL^*}$. We do so by replacing the LL with its \gls{kkt} conditions;
    \item Solving the bilevel problem with a \gls{bnb} algorithm adapted from Xu \textit{et al.} \cite{xu2014exact}. 
\end{enumerate}
The solution strategies for all other (single and bilevel) decision models (see Section \ref{sec: Other problems}) can be viewed as subroutines, as they only require step 1 and 2.

% We like to mention that most works addressing trilevel optimization problems in e.g. power system contexts apply Column-and-Constraint Generation methods \cite{xiang2018improved}\cite{gan2021tri}. This approach was pioneered by Zeng \textit{et al.} for trilevel problems arising in robust optimization \cite{zeng2013solving}. However, these methods can only be applied to trilevel problems in which $\mathcal{C}_{\mathrm{ML}}$ is independent of UL variables. The solution method from this paper thus targets a different set of trilevel problems. 
\subsection{Linearization}
\begin{figure}[!t]
\centering
\includegraphics[width=0.55 \linewidth]{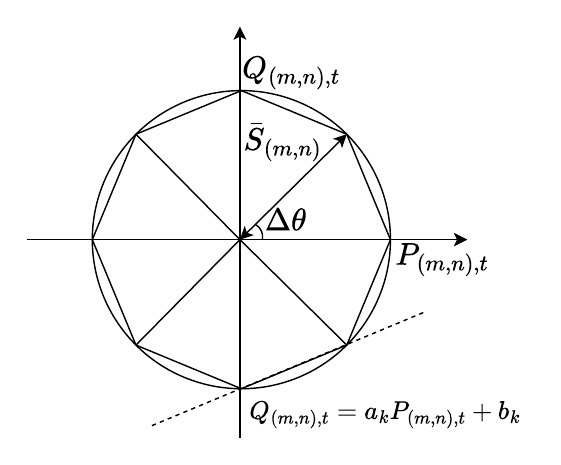}
\caption{Illustration of the convex restriction given by linear inequalities \eqref{eq: restriction} to approximate the circular feasibility region given by equation \eqref{eq: S ub} for some branch $(m, n) \in \Omega_B$ and some time step $t\in \Omega_T$}
\label{fig: restriction}
\end{figure}

The LL problem given by \eqref{eq: F LL}-\eqref{eq: ev soc 2} is convex, but not linear due to second-order cone constraints \eqref{eq: S ub}. In principle, the second and third step of this solution approach can be carried out with this constraint in place, but it greatly increases the solving times. We therefore introduce a convex restriction of the feasible space by introducing a set of linear constraints, as illustrated in \autoref{fig: restriction}. We divide the circle described by \eqref{eq: S ub} in the $P-Q$ plane into $K$ segments with angle $\Delta \theta$. We then derive the linear equations of the lines through points at an angle $k\Delta \theta$ and $(k+1)\Delta \theta$, where $k=0, \ldots ,K - 1$. If we take $K$ to be even, the expressions for the cuts for all branches $(m, n)\in \Omega_B$ and $t\in \Omega_T$ are given by:

\begin{equation}Q_{(m, n),t}
      \begin{cases}
     \leq a_k P_{(m, n),t} + b_k\bar{S}_{(m, n)} & \text{for } k < \frac{K}{2}, \\
     \geq a_k P_{(m, n),t} + b_k\bar{S}_{(m, n)} & \text{for } k \geq \frac{K}{2},
  \end{cases}
  \label{eq: restriction}
\end{equation}
where the coefficients are given by:
\begin{comment}
$$
a_k = \frac{\sin(n\Delta \theta)-\sin((n+1)\Delta \theta)}{\cos(n\Delta \theta) - \cos((n+1)\Delta \theta)}
$$
\end{comment}
$$
a_k = \cot\left(\left(k + \frac{1}{2}\right)\Delta \theta\right)
$$
and:
\begin{comment}
$$
b = \bar{S}_{(m, n),t}\frac{\cos(n\Delta \theta)\sin((n+1)\Delta \theta)-\cos((n+1)\Delta \theta)\sin(n\Delta \theta)}{\cos(n\Delta \theta) - \cos((n+1)\Delta \theta)}.
$$
\end{comment}
$$
b_k = \frac{\cos\left(\frac{1}{2}\Delta \theta\right)}{\sin\left((k+\frac{1}{2})\Delta \theta \right)}.
$$
This set of linear constraint presents a restriction of the feasible space of the \gls{dso}, potentially resulting in oversolving congestion with redispatch. By decreasing the angle resolution $\Delta \theta$ more accurate estimates of \eqref{eq: S ub} can be obtained. In this paper, we consider $\Delta \theta = \frac{1}{4}\pi$, like presented in \autoref{fig: restriction}.

\subsection{Removing the LL problem}
The LL problem described by \eqref{eq: F LL}-\eqref{eq: V2 lb ub}, \eqref{eq: aca 2}-\eqref{eq: ev soc 2}, \eqref{eq: restriction} is now a linear program. By ensuring that Slater's condition holds for our case study, the \gls{kkt} conditions of the LL are necessary and sufficient, and can replace the LL problem. The resulting stationarity, primal feasibility, dual feasibility, and \gls{cs} conditions can be added as constraints to the ML problem, resulting in a bilevel optimization problem consisting of the original UL and the new $\mathrm{LL^*}$ problem. This new $\mathrm{LL^*}$ \gls{mpec} has an objective function $F_{\mathrm{ML}}$ and its constraint set is the union of $\mathcal{C}_{\mathrm{ML}}$ and the \gls{kkt} conditions of the LL. These \gls{kkt} conditions are provided as Supplementary Material to this paper. The \gls{cs} conditions are linearized by the standard big-M method with values of $M=10^4$. Although the original ML constraints \eqref{eq: aca}-\eqref{eq: ev soc} were continuous and linear, applying the big-M method introduces binary variables to the constraint set of the new $\mathrm{LL^*}$, making it a \gls{milp}. 

\subsection{Brand-and-Bound Algorithm}
\label{sec: BnB}
Solving bilevel optimization problems with \gls{milp} lower levels are very challenging, as they cannot be reduced to an equivalent single-level formulation. Furthermore, not only are these problems NP-hard to solve, but so is checking whether some candidate solution is feasible \cite{kleinert2021survey}. Fortunately, our bilevel problem has the following three properties:
\begin{enumerate}
    \item All UL variables (i.e. $z^{\mathrm{ACA}}_{n, t}$) are discrete,
    \item All UL variables are bounded,
    \item No $\mathrm{LL^*}$ variables appear in $\mathcal{C}_{\mathrm{UL}}$.
\end{enumerate}
These properties imply that, in principle, we can apply the following enumeration approach: 1) list all possible candidate solutions of the UL problem $\hat{z}^{\mathrm{ACA}}_{n, t}$, 2) solve the $\mathrm{LL^*}$ \glspl{milp} parameterized by $\hat{z}^{\mathrm{ACA}}_{n, t}$, 3) list the resulting values of $F_{\mathrm{UL}}$, and 4) select the optimal solution of the bilevel problem from that list. 

Unfortunately, the number of possible candidate solutions $\hat{z}^{\mathrm{ACA}}_{n, t}$ is equal to $|\Omega_{\mathrm{FSP}}|\cdot|\Omega_{T}|!/(|\Omega_{T}| - B^{\mathrm{ACA}})!$, which would already result in solving 24.288 \glspl{milp} for a \gls{fsp} portfolio of 2 nodes and $B^{\mathrm{ACA}}=3$ for only a single sample $\omega$.

For this reason, we apply the exact \gls{bnb} algorithm developed by Xu \textit{et al.} for a specific set of bilevel mixed integer linear programming problems, satisfying three assumptions \cite{xu2014exact}. The first two assumptions correspond with properties 1 and 2 of our problem. The third assumption is that the coefficients of the UL variables in the constraints of the lower-level problem $\mathrm{LL^*}$ (i.e. the \textit{linking constraints}) are integers. For our problem, the linking constraints are equations \eqref{eq: aca}, \eqref{eq: aca 2}, and finally:
\begin{align}
\bar{p}_{n} + (\bar{p}^{\mathrm{ACA}}_{n} - \bar{p}_{n})z^{\mathrm{ACA}}_{n, t} \leq p^{\mathrm{EV}}_{n, t} + M z^{\mathrm{CS}}_{n, t} \label{eq: linking M} && n \in \Omega_N, t \in \Omega_T, 
\end{align}
which is a \gls{cs} constraint of \eqref{eq: aca 2}, linearized with the big-M method with auxiliary binary variables $z^{\mathrm{CS}}_{n, t}$. The coefficients of UL variables in the linking constraints are thus integers if and only if the values of $(\bar{p}^{\mathrm{ACA}}_{n} - \bar{p}_{n})$ are integers. For our case study (Section \ref{sec: Results}), we thus have to select the values of $\bar{p}^{\mathrm{ACA}}_{n}$ in the \gls{aca} such that this holds.

For details on the \gls{bnb} method, we refer to \cite{xu2014exact}. Its strength lies in branching on linking constraints, which all have a very similar structure in our problem. This results in many equivalent \gls{bnb} nodes at every branching step, which can be filtered. Furthermore, due to the simplicity of the constraints in $\mathcal{C}_{\mathrm{UL}}$, many of the infeasible nodes could be removed before evaluation. We therefore added an extra step to the algorithm of Xu \textit{et al.} to prune away all redundant or infeasible nodes during branching. This resulted in only a very moderate number of evaluated nodes during the solving process (see the results in Section \ref{sec: Results}), making this method very suitable for our problem. The evaluation of a \gls{bnb} node involves solving a \gls{milp} for calculating the bound, and solving the $\mathrm{LL^*}$ \gls{milp}. Both were solved with Gurobi \cite{gurobi} with a MIPGap of 0.001\%.

%inefficient on paper, because the branching method creates as many new nodes in every branching step as there are linking constraints in the bilevel problem. 
%For our problem, this number of nodes grows with the size of $\Omega_{\mathrm{FSP}}$ and $\Omega_{T}$, but also with $\Omega_S$ if we would have taken a more risk-aware approach. 
% However, for our problem, the 
   
\section{Results}
\label{sec: Results}
\begin{figure}[!t]
\centering
\includegraphics[width=0.60\linewidth]{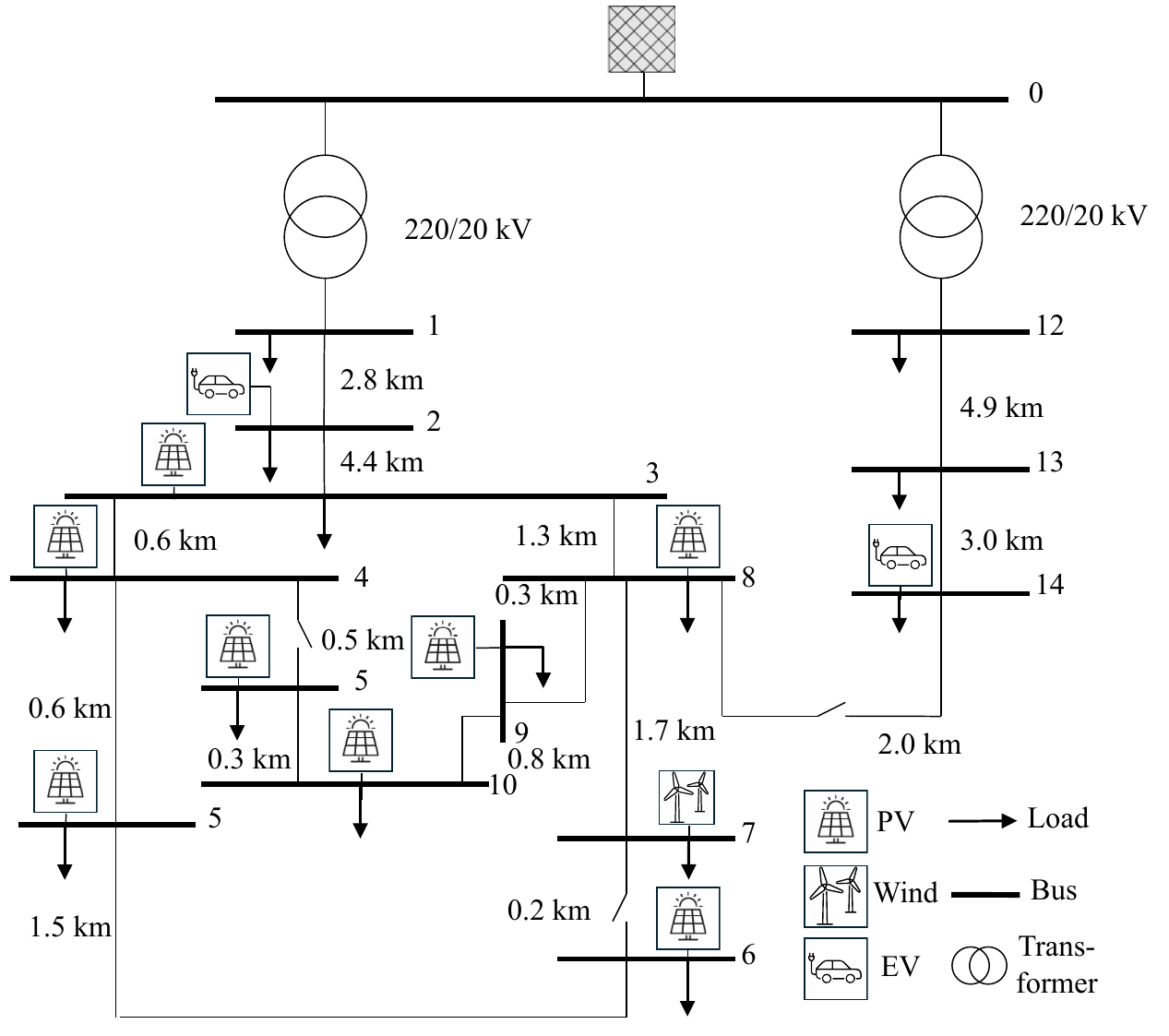}
\caption{Schematic representation of the CIGRE network with solar and wind generation under radial operation, adapted from \cite{rudion2006design}. The simulated \gls{fsp} controls two \gls{ev} fleets at buses 2 and 14.}
\label{fig: network}
\end{figure}

\begin{figure}[!t]
\centering
\includegraphics[width=1.0\linewidth]{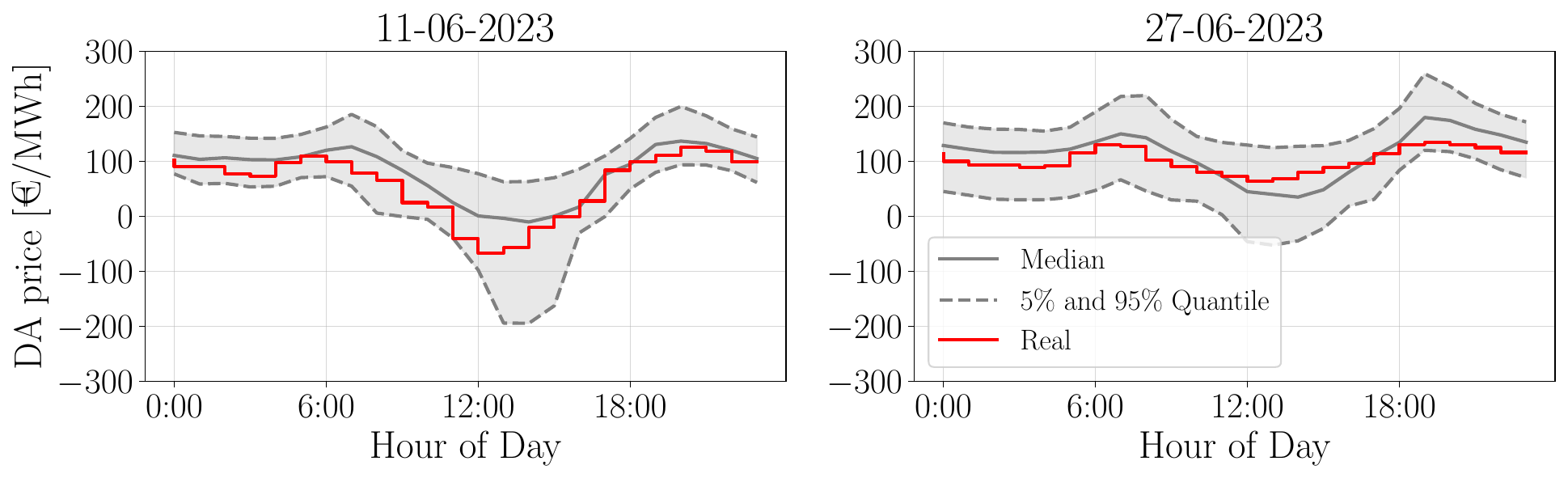}
\caption{The day-ahead market prices for the two simulated days. The actual data for the Dutch market is presented in red, and the median and 5\%-95\% quantile range are presented for the scenarios.}
\label{fig: da prices}
\end{figure}

The results are obtained on the radially operated CIGRE medium-voltage distribution network with photovoltaics and wind, presented in \autoref{fig: network}, and for the parameters presented in \autoref{tab: parameters}. It is assumed the \gls{fsp} controls two \gls{ev} fleets, which are at buses $2$ and $14$. We assume a constant power factor for the \gls{ev} charging of $1.0$. We will simulate two different days, for which the actual \gls{da} prices and scenarios are presented in \autoref{fig: da prices}.

\begin{table}[h!]
  \centering
  \vspace{-0.5cm}
  \caption{Model Parameters}
  \label{tab: parameters}
  \begin{tabular}{l l|l|l|l}
    \hline  
    $\pi^{\mathrm{RC}}$ & [€/MWh]& 100.00  & $B^{\mathrm{ACA}}$ [-]& 3\\ 
    $\Delta \theta$ & rad & $\pi/4$ & $\sigma$ \hspace{15pt} [-] & 0.01 \\ 
    \hline
    & & Fleet Size & $\bar{p}$ [MW] & $\bar{p}^{\mathrm{ACA}}$ [MW] \\ % Header for the second section
    \hline
    Node 2 & & 10.000 & 12.0 & 2.0\\ 
    Node 14 & & 5.000 & 6.0 &  2.0\\
    \hline
  \end{tabular}
    \vspace{-0.4cm}
\end{table}

\subsection{Gaming versus No Gaming Strategy}
\label{sec: Results A}
\begin{figure}[!b]
\centering
\vspace{-0.5cm}
\includegraphics[width=0.75\linewidth]{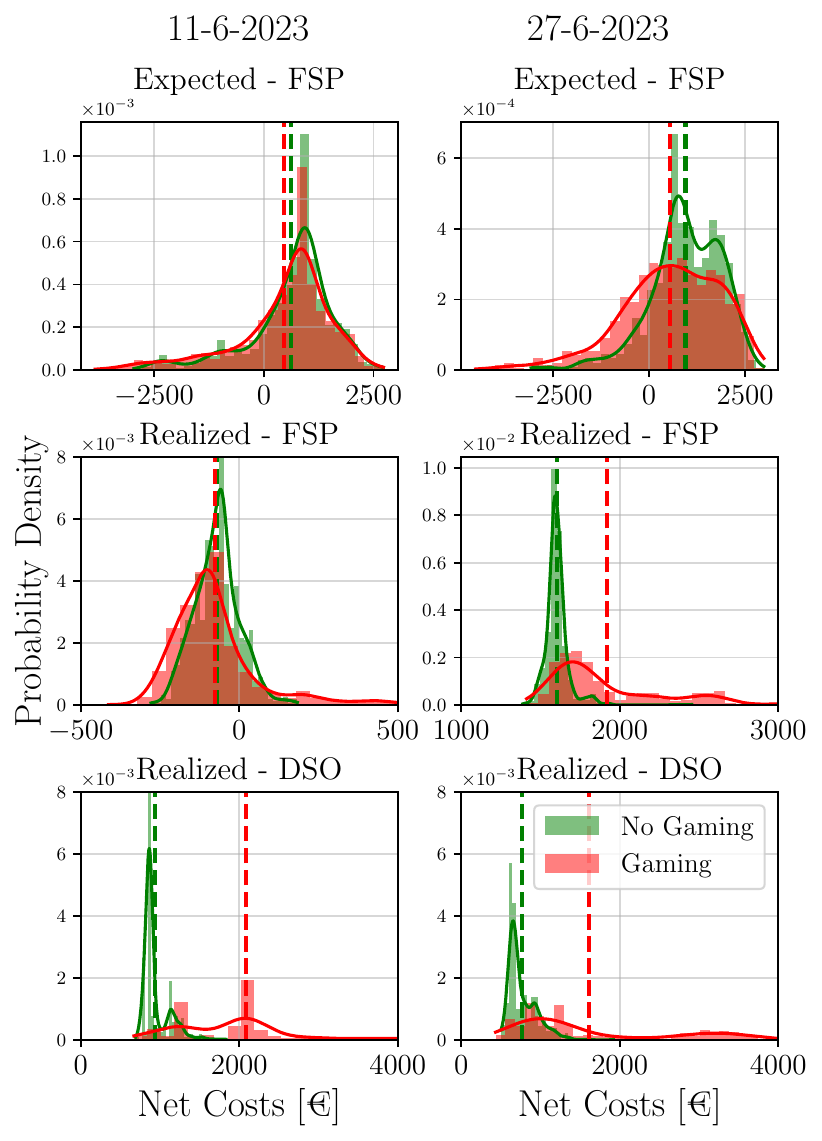}
\caption{Normalized histogram and Kernel Density Estimations of the (expected) net costs of the \gls{fsp} and the \gls{dso} for the two simulated days. The net costs for \gls{fsp} consist of the costs on the \gls{da} market minus the revenues from redispatch. The net costs from the \gls{dso} are just the redispatch costs. The means of the distributions are represented by the dashed lines.}
\label{fig: 6 histograms}
\end{figure}

We first compare the \textit{No Gaming} and \textit{Gaming} strategies of the \gls{fsp} for the case the \gls{dso} does not exercise the \glspl{aca}. The simulation results are thus obtained by solving an Attacker-Defender game, consisting of the second and third step of the game presented in \autoref{fig: actors}. \autoref{fig: 6 histograms} presents the normalized histograms and \glspl{kde} of the expected and realized net costs of the \gls{fsp} and the \gls{dso} for the two simulated days. The results in green presents the case where the \gls{fsp} only minimizes \gls{da} costs (\textit{No Gaming}), whereas the results in red correspond to the case where the \gls{fsp} constructs its baseline while anticipating redispatch revenues (\textit{Gaming}). The net costs for the \gls{fsp}, given by equation \eqref{eq: F ML}, consist of the \gls{da} market costs minus the revenues from redispatch. The \gls{dso}'s net costs $F_{\mathrm{UL}}$ consist solely of redispatch costs. Expected costs are calculated in the day-ahead stage based on the scenarios, while realized costs are calculated after the \gls{dso} applied redispatch close to delivery, using actual load and price realizations.

The expected costs of the \gls{fsp} presented in the first row of \autoref{fig: 6 histograms} illustrate how much the \gls{fsp} expects to benefit from applying increase-decrease gaming. We observe that on 11-6, the \gls{fsp} expects to reduce its mean costs by only €$166.09$ compared to €$403.80$ on 27-6 by applying the \textit{Gaming} strategy. The reason for this difference is that \gls{da} prices on 11-6 shown in \autoref{fig: da prices}, were expected to be more volatile and even negative. As a result, charging the \gls{ev} fleets under minimum \gls{da} prices during the afternoon results in the lowest costs for both \gls{fsp} strategies. On 27-6, the \gls{da} prices were expected to be less volatile. This is why the \gls{fsp} more often saw an opportunity to increase its baseline during times of slightly higher \gls{da} prices to benefit from more redispatch revenues later. On 27-6, the \gls{fsp} was thus more actively gaming its baseline for higher redispatch revenues than it was doing on 11-6.

However, when inspecting the realized costs for the \gls{fsp}, it can be observed that this strategy does not necessarily benefit the \gls{fsp} on 27-6. As a matter of fact, choosing this gaming strategy under uncertain loads and \gls{da} prices increased the mean costs of the \gls{fsp} by $16.4\%$. Further analysis shows that the effect of load uncertainty was negligible in this case, as the results do not change significantly for values of $\sigma \in [0.0, 0.05]$ (data not shown). Rather, the main reason is the \gls{da} price uncertainty, as the actual \gls{da} prices for 27-6 are very close to the redispatch compensation of $\pi^{\mathrm{RC}} = $ €$100 \mathrm{/MWh}$ (see \autoref{fig: da prices}). Hence, in hindsight, the extra redispatch revenues from increase-decrease gaming were smaller than expected. The \gls{fsp} expected to increase its \gls{da} costs to get higher redispatch revenues in return, but the extra revenues were not there due to small price differences between the two markets. This result shows the importance of price risk on the success of the gaming strategy of the \gls{fsp}.

\begin{figure}[!b]
\centering
\vspace{-0.5cm}
\includegraphics[width=0.7\linewidth]{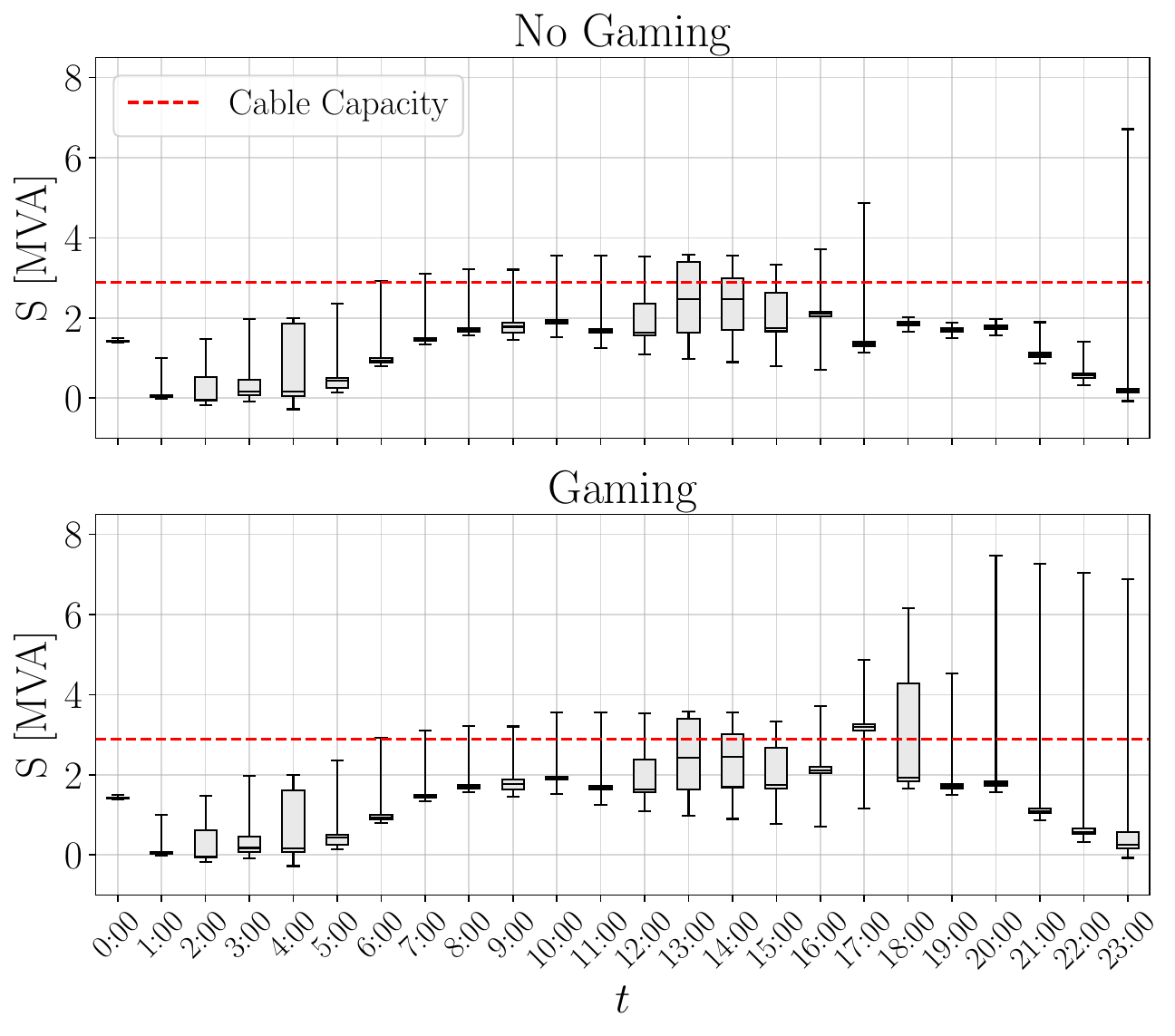}
\vspace{-0.3cm}
\caption{The expected power flow between bus 1 and 2 according to the baselines of the \gls{fsp} for the two strategies. In the first plot, the \gls{fsp} constructed the baseline based solely on predicted \gls{da} costs. For the second plot, the \gls{fsp} was able to anticipate redispatch income, resulting in increase-decrease gaming.}
\label{fig: line loadings}
\end{figure}

When analyzing the redispatch costs of \gls{dso} in \autoref{fig: 6 histograms}, it is clear that for both days the costs increase significantly if the \gls{fsp} adopts the \textit{Gaming} strategy. Also, we can see that on 11-6, the added costs for the \gls{dso} are relatively large compared to the added revenues of the \gls{fsp}. \autoref{fig: line loadings} shows the expected loading through the cable between bus 1 and 2 for the two strategies. There we see that some congestion is expected between 14:00 and 16:00 for both strategies due to charging at low \gls{da} prices. However, for the \textit{Gaming} strategy, we see the \gls{fsp} increasing its baseline for the \gls{ev} fleet at the end of the day, particularly between 17:00 and 19:00. At these times, the other residental loads in the network are high, and many cars of the \gls{ev} fleet returned home, making them available to the \gls{fsp} for increase-decrease gaming. These are ideal circumstances for the \gls{fsp} to increase its baseline, pushing the cable in a congested state.

\begin{figure}[!t]
\centering
\vspace{-0.25cm}
\includegraphics[width=0.9\linewidth]{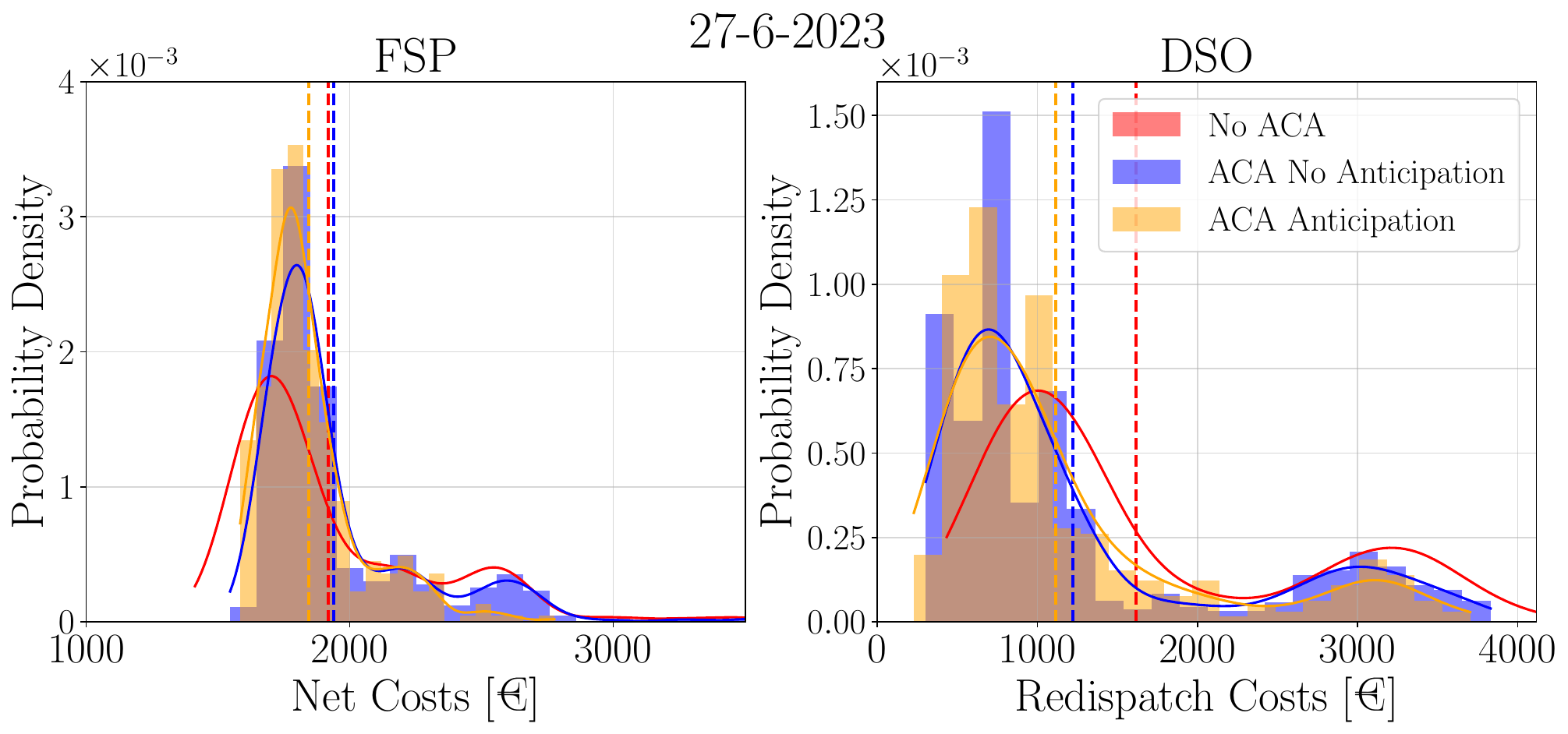}
\caption{Normalized histogram and Kernel Density Estimations of the realized costs of the \gls{fsp} and the \gls{dso} for day 27-6-2023. For all presented strategies the \gls{fsp} applies the gaming strategy, and the \gls{dso} takes the following strategies: not applying the \gls{aca} (red), applying the \gls{aca} using the \textit{ACA No Anticipation} strategy (blue), or applying the \gls{aca} using the \textit{ACA Anticipation} strategy (orange). The histogram for the first case is not presented, as it is already shown in \autoref{fig: 6 histograms}.}
\vspace{-0.3cm}
\label{fig: ACA hist}
\end{figure}
\subsection{Applying ACA with/without Anticipation}
We now investigate the effect of the \gls{dso} exercising \glspl{aca}, given the \gls{fsp} adopts the \textit{Gaming} strategy. \autoref{fig: ACA hist} presents the realized net costs and redispatch costs on 27-6-2023 for the \gls{fsp} and \gls{dso}, respectively. Results for both the \textit{ACA No Anticipation} and \textit{ACA Anticipation} strategies of the \gls{dso} are shown. As a reference, we also show the case for which the \gls{dso} does not apply the \gls{aca} at all, being the same data as presented under the \textit{Gaming} strategy in \autoref{fig: 6 histograms}. We observe that applying the \gls{aca} can reduce the mean redispatch costs of the \gls{dso} by $24.4\%$ and $31.0\%$ for the \textit{ACA No Anticipation} and \textit{ACA Anticipation} strategies, respectively. We especially see for the latter that the number of scenarios with very high redispatch costs have reduced. By design, the \textit{ACA No Anticipation} strategy only applied the \gls{aca} between 14:00-16:00 to prevent anticipated congestion caused by simultaneous charging on low \gls{da} prices (data not shown). The costs for the \gls{fsp} are impacted significantly less than the \gls{dso}'s, with close to 0\% change for the \textit{ACA No Anticipation} strategy, and a decrease of $1.97\%$ for the \textit{ACA Anticipation} strategy. The decrease comes from the \gls{dso} preventing the \gls{fsp} from making unfavorable gaming decisions, resulting in a win-win situation for both parties.

\begin{figure}[!t]
\centering
\includegraphics[width=0.55\linewidth]{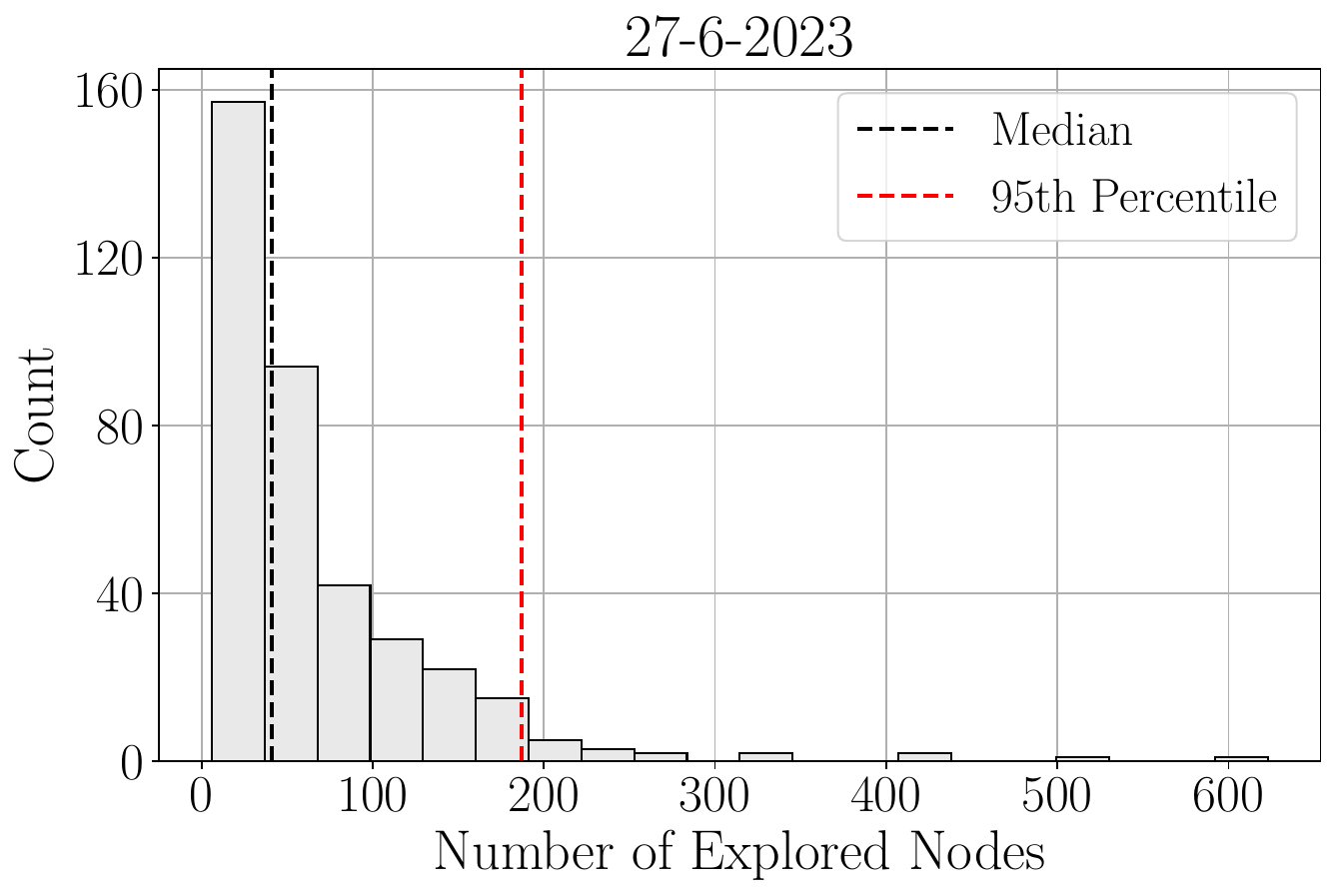}
\caption{Histogram of the number of explored nodes in the branch-and-bound algorithm to solve the trilevel optimization problem from the case study. Every count corresponds with a scenario for which the problem was solved.}
\label{fig: Node count hist}
\end{figure}
\autoref{fig: Node count hist} presents the number of explored nodes in the \gls{bnb} algorithm for the trilevel problem solved for the \textit{ACA Anticipation} strategy. We observe that for most scenarios, only $10^1-10^2$ nodes needed to be explored, with a median of 41 nodes. For most scenarios, the solution strategy presented in Section \ref{sec: Solution Method} solved the trilevel optimization problem with limited computational effort. However, we also observe that for some instances significantly more nodes were explored. \autoref{tab: B ACA} presents the median and 95th percentile of the number of explored nodes for various values of the daily \gls{aca} budget $B^{\mathrm{ACA}}$. We also provide the number of nodes needed to solve the problem using the brute-force enumeration approach discussed in Section \ref{sec: BnB}, and we provide the realized cost decrease for the \gls{dso}. This decrease is given in relative terms, compared to the situation in which the \gls{dso} does not exercise any \glspl{aca}. The cost decrease of the \gls{fsp} was very insensitive to the values of $B^{\mathrm{ACA}}$ (at most $3.7\%$) and is therefore not shown.

We observe that though the combinatorial complexity of the problem grows quickly in $B^{\mathrm{ACA}}$, that the median number of explored nodes grows only slowly (more or less linearly). This shows that our solution method solves the trilevel problem efficiently for the majority of scenarios. For the 95th percentile, the increase is faster, approximating an exponential increase in $B^{\mathrm{ACA}}$. When looking at the cost saving for the \gls{dso}, we see that the cost decrease grows for larger daily budgets of \gls{aca} activation. Furthermore, we see that the \textit{ACA Anticipation} strategy becomes increasingly valuable when more \gls{aca} activations can be applied, whereas the strategy without anticipation stagnates around $25\%$. 

\begin{table}[h!]
  \centering
  \caption{Number of explored nodes and relative redispatch cost reduction for various ACA daily budgets}
  \label{tab: B ACA}
   \scriptsize
  \begin{tabular}{l|l|l|l|l|l|l}
     $B^{\mathrm{ACA}}$ & 0 & 1 & 2 & 3 & 4 & 5\\ 
    \hline 
     Median & 2 & 10 & 25 & 41 & 50 & 72\\ 
    95th percentile & 2 & 17 & 80 & 187 & 421 & 765\\
    \makecell[l]{Max Possible \\ Nodes} & 2 & 48 & 1.104 & 24.288 & 510.048 & 10.200.960\\
    % Anticipation CSP & 1919.73 & 1873.95 & 1920.84 & 1882.62 & 1871.08 & 1847.83\\
    % Anticipation DSO [\%] & 1614.95 & 1332.89 & 1296.17 & 1115.07 & 1009.73 & 918.66\\
    % No Anticipation CSP & 1919.73 & 1927.95 & 1920.63 & 1919.92 & 1917.43 & 1915.94\\
    % No Anticipation DSO [\%] & 1614.95 & 1501.67 & 1297.40 & 1220.73 & 1194.00 & 1171.00\\
    \makecell[l]{ACA No [\%] \\Anticipation } & 0.0 & 11.2 & 19.7 & 24.4 & 26.1 & 27.5\\
    \makecell[l]{ACA [\%]\\ Anticipation } & 0.0 & 17.5 & 24.8 & 31.0 & 37.5 & 43.1\\
    \hline
  \end{tabular}
\end{table}
\vspace{-0.3cm}

\section{Conclusion}
\label{sec: Conclusion}
This paper presented the first quantitative study on how \glspl{dso} can use \glspl{aca} to prevent increase-decrease gaming in redispatch markets. We develop a defender-attacker-defender game to investigate how \glspl{dso} can strategically use \glspl{aca} to prevent \glspl{fsp} from increasing redispatch income by manipulating their baselines. The investigated model considers the \gls{aca} and redispatch market as currently available in the Netherlands. In this game, the \gls{dso} solves a trilevel optimization model, for which we develop a solution strategy based on an adapted branch-and-bound algorithm. We demonstrate that for most scenarios, the solution strategy efficiently solves the problem without exploring many nodes. We include both load and price uncertainty in our analysis and show that day-ahead price uncertainty can play a large role in the effectiveness of the increase-decrease gaming strategy. Furthermore, we demonstrate that using \glspl{aca} can significantly reduce redispatch costs for the \gls{dso}, while not decreasing profits for \glspl{fsp} at the same time. It was shown that the effectiveness of applying \glspl{aca} for this purpose depends on the number of times the \gls{dso} can invoke the \gls{aca} on a daily basis, and whether the \gls{dso} can strategically anticipate the gaming strategy of the \gls{fsp}. The latter can be difficult to achieve in practice, but even without this condition, redispatch savings of around 25\% were observed in our case study. 

% Future research could focus on the influence of risk preferences of \glspl{fsp} and \glspl{dso} on increase-decrease gaming and its prevention with \glspl{aca}, respectively. Furthermore, other \gls{aca} designs than the ATR85 from the Netherlands could be investigated. 
% Finally, relaxing the assumption of complete information could be an interesting research direction to bring the developed framework closer to a practical tool for \gls

\begin{comment}
\section*{Acknowledgments}
This should be a simple paragraph before the References to thank those individuals and institutions who have supported your work on this article.
\end{comment}
% \input{Appendix/Appendix}

 

\bibliographystyle{IEEEtran} 
\bibliography{bibliography}

\clearpage   % ensure bibliography ends properly
\onecolumn   % switch to one column cleanly
\appendices  % (optional) to make section numbering consistent

\begin{comment}

\twocolumn[\section{Supplementary Material - KKT conditions of the LL problem}] % keeps section title wide
\onecolumn  % <-- switch to one column for wide appendix content
    
\end{comment}
\section*{Supplementary Material — KKT Conditions of the LL Problem}
We now present the KKT conditions of the linearized LL problem, required in Section \ref{sec: Solution Method}. The linearized LL is given by:
\begin{equation}
% \vspace{-2cm}
F_{\mathrm{LL}}(E^{\mathrm{RC}}_{n, t}) = \sum_{n \in \Omega_{\mathrm{FSP}}} \sum_{t\in\Omega_T} \Delta \pi_{t, \omega}^{\mathrm{RC}} E^{\mathrm{RC}}_{n, t} \tag{\ref{eq: F LL}},
% \vspace{-2cm}
\end{equation}
\begingroup
\allowdisplaybreaks
\begin{align}  
     & E^{\mathrm{RC}}_{n, t} \geq 0, \tag{\ref{eq: redispatch 0}}  && :\ubar{\mu}_{n, t}^{\mathrm{RC}} && n \in \Omega_{\mathrm{FSP}}, t \in \Omega_T, \\ 
     & p^{\mathrm{EV}}_{n, t}\Delta t = p^{\mathrm{DA}}_{n, t}\Delta t - E^{\mathrm{RC}}_{n, t}, \tag{\ref{eq: redispatch}}  && :\lambda_{n, t}^{\mathrm{RC}} && n \in \Omega_{\mathrm{FSP}}, t \in \Omega_T, \\ 
     & p_{n, t} = p^{\mathrm{IF}}_{n, t, \omega} + \delta^{\mathrm{EV}}_{n} p^{\mathrm{EV}}_{n, t}, \tag{\ref{eq: p balance}} && :\lambda_{n, t}^{p} && n \in \Omega_N, t \in \Omega_T, \\
      & q_{n, t} = \lambda^{\mathrm{IF}}_{n} p^{\mathrm{IF}}_{n, t, \omega} + \delta^{\mathrm{EV}}_{n} \lambda^{\mathrm{EV}} p^{\mathrm{EV}}_{n, t}, \tag{\ref{eq: q balance}} && :\lambda_{n, t}^{q} && n \in \Omega_N, t \in \Omega_T, \\
     & \sum_{(m, n)} P_{(m, n), t} - \sum_{(n, o)} P_{(n, o), t} = p_{n, t}, \tag{\ref{eq: P balance}} && :\lambda_{n, t}^{P} && n \in \Omega_N, t \in \Omega_T,\\
      & \sum_{(m, n)} Q_{(m, n), t} - \sum_{(n, o)} Q_{(n, o), t} = q_{n, t}, \tag{\ref{eq: Q balance}} && :\lambda_{n, t}^{Q} && n \in \Omega_N, t \in \Omega_T,\\
    & V_{n, t}^{\mathrm{sqr}} = V_{m, t}^{\mathrm{sqr}} + 2\big(R_{(m, n)} P_{(m, n),t}\nonumber \\
&\quad\quad + X_{(m, n)}Q_{(m, n),t}\big), \tag{\ref{eq: V2}} && :\lambda_{(m, n), t}^{V} && (m, n) \in \Omega_B, t \in \Omega_T, \\
      & \ubar{V}^{\mathrm{sqr}} \leq V_{n, t}^{\mathrm{sqr}}\leq \bar{V}^{\mathrm{sqr}}, \tag{\ref{eq: V2 lb ub}} && :\ubar{\mu}_{n, t}^{V},\bar{\mu}_{n, t}^{V} && n \in \Omega_N, t \in \Omega_T, \\
      & 0 \leq p^{\mathrm{EV}}_{n, t}\leq \bar{p}_{n} + (\bar{p}^{\mathrm{ACA}}_{n} - \bar{p}_{n})z^{\mathrm{ACA}}_{n, t},\tag{\ref{eq: aca 2}} && :\ubar{\mu}_{n, t}^{\mathrm{ACA}},\bar{\mu}_{n, t}^{\mathrm{ACA}} && n \in \Omega_{\mathrm{FSP}}, t \in \Omega_T, \\
     & 0 \leq  p^{\mathrm{EV}}_{n, t}\leq \bar{p}^{\mathrm{EV}}_{n, t}, \tag{\ref{eq: ev pmax 2}} && :\ubar{\mu}_{n, t}^{p},\bar{\mu}_{n, t}^{p} && n \in \Omega_{\mathrm{FSP}}, t \in \Omega_T, \\
     &\bar{E}_{n, t} \leq \sum_{\tau \leq t}  p^{\mathrm{EV}}_{n, \tau} \Delta t\leq \ubar{E}_{n, t}, \tag{\ref{eq: ev soc 2}} && :\ubar{\mu}_{n, t}^{E},\bar{\mu}_{n, t}^{E}  && n \in \Omega_{\mathrm{FSP}}, t \in \Omega_T,\\
      & Q_{(m, n),t}  \geq a_k P_{(m, n),t} + b_k\bar{S}_{(m, n)}\tag{\ref{eq: restriction}} && :\ubar{\mu}_{(m, n), t}&& (m, n) \in \Omega_B, t \in \Omega_T, k \geq \frac{K}{2},\\
     & Q_{(m, n),t}  \leq a_k P_{(m, n),t} + b_k\bar{S}_{(m, n)}\tag{\ref{eq: restriction}} && :\bar{\mu}_{(m, n), t}&& (m, n) \in \Omega_B, t \in \Omega_T, k < \frac{K}{2},
\end{align}
\endgroup
where we included all the dual variables of the constraints. All dual variables denoted by $\lambda$ correspond with equality constraints. All dual variables denoted by $\mu$ correspond with inequality constraints and $\mu \geq 0$ (dual feasibility).

With every inequality constraint comes a complementary slackness (CS) condition. If we take constraint \eqref{eq: aca 2} as an example, this results in the constraints:
\begin{align}
    \ubar{\mu}_{n, t}^{\mathrm{ACA}}\cdot - p^{\mathrm{EV}}_{n, t} = 0 && n \in \Omega_{\mathrm{FSP}}, t \in \Omega_T,\\
    \bar{\mu}_{n, t}^{\mathrm{ACA}} (p^{\mathrm{EV}}_{n, t} - \bar{p}_{n} - (\bar{p}^{\mathrm{ACA}}_{n} - \bar{p}_{n})z^{\mathrm{ACA}}_{n, t})) = 0 && n \in \Omega_{\mathrm{FSP}}, t \in \Omega_T.
\end{align}
Applying the big-M method, these equalities can be written as:
\begingroup
\allowdisplaybreaks
\begin{align}
     \ubar{\mu}_{n, t}^{\mathrm{ACA}} &\leq (1 - \ubar{z}_{n, t}^{\mathrm{CS}})   M  && n \in \Omega_{\mathrm{FSP}}, t \in \Omega_T,\\
     p^{\mathrm{EV}}_{n, t} &\leq  \ubar{z}_{n, t}^{\mathrm{CS}}  M && n \in \Omega_{\mathrm{FSP}}, t \in \Omega_T,\\
     \bar{\mu}_{n, t}^{\mathrm{ACA}} &\leq (1 - \bar{z}_{n, t}^{\mathrm{CS}})M  && n \in \Omega_{\mathrm{FSP}}, t \in \Omega_T,\\
     \bar{p}_{n} + (\bar{p}^{\mathrm{ACA}}_{n} - \bar{p}_{n})z^{\mathrm{ACA}}_{n, t}) - p^{\mathrm{EV}}_{n, t}  & \leq  \bar{z}_{n, t}^{\mathrm{CS}}M \label{eq: linking M A} && n \in \Omega_{\mathrm{FSP}}, t \in \Omega_T,\\
     \ubar{z}_{n, t}^{\mathrm{CS}} & \in \{0, 1\} && n \in \Omega_{\mathrm{FSP}}, t \in \Omega_T,\\
     \bar{z}_{n, t}^{\mathrm{CS}} & \in \{0, 1\} && n \in \Omega_{\mathrm{FSP}}, t \in \Omega_T,
\end{align}
\endgroup
for a large constant $M$. Note that equation \eqref{eq: linking M A} was also mentioned as equation \eqref{eq: linking M} in Section \ref{sec: Solution Method}.

Furthermore, the stationarity conditions are given by:
\begingroup
\allowdisplaybreaks
\begin{empheq}{alignat=2}
     \Delta \pi_{t, \omega}^{\mathrm{RC}} + \lambda_{n, t}^{\mathrm{RC}} -  \ubar{\mu}_{n, t}^{\mathrm{RC}}= 0  & :E^{\mathrm{RC}}_{n, t} \quad  && n \in \Omega_{\mathrm{FSP}}, t \in \Omega_T, \\ 
\begin{aligned}\lambda_{n, t}^{\mathrm{RC}}\Delta t -  \delta^{\mathrm{EV}}_{n}\lambda_{n, t}^p -  \delta^{\mathrm{EV}}_{n}\lambda^{\mathrm{EV}}\lambda_{n, t}^p + \\ \bar{\mu}_{n, t}^{p} - \ubar{\mu}_{n, t}^{p} +  \bar{\mu}_{n, t}^{\mathrm{ACA}} - \ubar{\mu}_{n, t}^{\mathrm{ACA}} + \sum_{\tau \geq t}(\bar{\mu}_{n, \tau}^{E} - \ubar{\mu}_{n, \tau}^{E})\Delta t  = 0\end{aligned} &\quad :p^{\mathrm{EV}}_{n, t} && n \in \Omega_{\mathrm{FSP}}, t \in \Omega_T, \\ 
\lambda_{n, t}^{p} -  \lambda_{n, t}^{P}=0, &\quad:p_{n, t} && n \in \Omega_N, t \in \Omega_T, \\ 
\lambda_{n, t}^{q} -  \lambda_{n, t}^{Q}  = 0  &\quad:q_{n, t} && n \in \Omega_N, t \in \Omega_T, \\ 
\begin{aligned}\lambda_{m, t}^{P} -  \lambda_{n, t}^{P} - 2 R_{(m, n)}\lambda^V_{(m, n), t} +  a_k \ubar{\mu}_{(m, n), t} = 0\end{aligned}  &\quad:P_{(m, n), t} && n \in \Omega_N, t \in \Omega_T, k \geq \frac{K}{2},\\ 
\begin{aligned}\lambda_{m, t}^{P} -  \lambda_{n, t}^{P} - 2 R_{(m, n)}\lambda^V_{(m, n), t} -  a_k \bar{\mu}_{(m, n), t} = 0\end{aligned}  &\quad:P_{(m, n), t} && n \in \Omega_N, t \in \Omega_T, k < \frac{K}{2},\\ 
\begin{aligned}\lambda_{m, t}^{Q} -  \lambda_{n, t}^{Q} - 2 X_{(m, n)}\lambda^V_{(m, n), t} - \ubar{\mu}_{(m, n), t}= 0\end{aligned}  &\quad :Q_{(m, n), t}\quad && (m, n) \in \Omega_B, t \in \Omega_T, k \geq \frac{K}{2},\\ 
\begin{aligned}\lambda_{m, t}^{Q} -  \lambda_{n, t}^{Q} - 2 X_{(m, n)}\lambda^V_{(m, n), t} + \bar{\mu}_{(m, n), t}= 0\end{aligned}  &\quad:Q_{(m, n), t} && (m, n) \in \Omega_B, t \in \Omega_T, k < \frac{K}{2},\\ 
\begin{aligned} \sum_{(m, n)} \lambda_{(m, n), t}^V -  \sum_{(n, o)} \lambda_{(n, o), t}^V +   \bar{\mu}_{n, t}^{V} - \ubar{\mu}_{n, t}^{V} = 0    
\end{aligned}  &\quad:V^{\mathrm{sqr}}_{n, t} && n \in \Omega_{N}, t \in \Omega_T.
\end{empheq}
\endgroup

 %\newpage
\begin{comment}

\section{Biography Section}
If you have an EPS/PDF photo (graphicx package needed), extra braces are
 needed around the contents of the optional argument to biography to prevent
 the LaTeX parser from getting confused when it sees the complicated
 $\backslash${\tt{includegraphics}} command within an optional argument. (You can create
 your own custom macro containing the $\backslash${\tt{includegraphics}} command to make things
 simpler here.)
      
\vspace{11pt}
\end{comment}

% \end{IEEEbiography}


% Use $\backslash${\tt{begin\{IEEEbiographynophoto\}}} and the author name as the argument followed by the % biography text.

\vfill

\end{document}